\documentclass[onecolumn]{aastex61}
\usepackage{amsmath}
\usepackage{verbatim}
\usepackage{graphicx}	
\usepackage{booktabs}
\usepackage{color}
\usepackage{lmodern}
\usepackage{epstopdf}
\usepackage{physics}
\usepackage{rotating}

\newcommand{\ix}[2]{#1_{\mathrm{#2}}}

\received{July 17, 2018}
\revised{October 22, 2018}
\accepted{November 3, 2018}

\submitjournal{ApJ}

\shorttitle{RHD Simulations of Low-Mass Collapse}
\shortauthors{Stamer \& Inutsuka}

\begin{document}

\title{Radiation Hydrodynamics Simulations of Spherical Protostellar Collapse for Very Low Mass Objects}

\author{Torsten Stamer}
\affil{Nagoya University \\
Furo-cho, Chikusa-ku, Nagoya, Japan}
\email{stamer.torsten@a.mbox.nagoya-u.ac.jp}

\author{Shu-ichiro Inutsuka}
\affil{Nagoya University \\
Furo-cho, Chikusa-ku, Nagoya, Japan}
\email{inutsuka@nagoya-u.jp}

\begin{abstract}

We perform radiation hydrodynamical simulations of protostellar collapse in spherical symmetry, with a special focus on very low-mass objects, i.e. brown dwarfs and sub-brown dwarfs. The inclusion of a realistic equation of state that includes the effect of hydrogen dissociation allows for a modeling of the complete process from the beginning of the collapse until the formation of the protostar. We solve the frequency-dependent radiative transfer equation without any diffusion approximation, using realistic dust and gas opacities.

Our results show that the properties of the protostar are essentially independent of the initial conditions, which had previously only been confirmed for higher mass ranges. For very low mass initial conditions, however, we find that the first core phase of the collapse shows some significant differences in the time evolution, with the first core lifetime increasing dramatically because of the reduced accretion rate from the surrounding envelope. We consider the observational implications of this. We also investigate the opposite case of a collapse without any first core phase, which may occur for very unstable initial conditions.

In the appendix, we describe a severe numerical problem that causes an unphysical expansion after the formation of the protostar, which may affect other attempts at similar calculations of self-gravitational collapse. We explain the origin of the unphysical behavior and present a solution that can be used in similar investigations.

\end{abstract}


\section{Introduction}
Stars form in molecular clouds - regions of the interstellar medium which are very dense ($> \mathrm{10^2 / cm^3}$) and cold (10 K), allowing for the hydrogen gas of which they consist to be present in molecular form ($\mathrm{H_2}$). Within these clouds, there exist regions of especially high density, usually called ``prestellar cores'' (or ``protostellar cores'' if a protostar is present within). It is believed that star formation occurs where such cloud cores become massive enough to collapse under their own gravity in a process called ``protostellar collapse'', the numerical simulation of which is the subject of this paper.

How these cores form in the first place is still a matter of active research. It has been firmly established by the \textit{Herschel} survey (e.g. \citet{Andre2014}) that the internal structure of molecular clouds is dominated by filaments, and that the majority of prestellar and protostellar cores in star-forming regions are found within them. Theoretical consideration of an isothermal, self-gravitating cylinder (\citet{inu1992,inu1997}) shows that filaments are gravitationally stable only if their line mass (i.e. mass per length) does not exceed a certain critical value $\ix{M}{line,crit} \approx 16 \ix{M}{\odot} \mathrm{pc^{-1}} \times (\ix{T}{gas}/\mathrm{10K})$. Intersections between different filaments may create the regions most conducive to star formation, and especially star cluster formation, due to localized higher density (\citet{Myers2009,Myers2011,Schneider2012}).
We may therefore state that star formation occurs in regions of overdense, radially collapsing filaments. \citet{inu1992,inu1997} studied the stability of self-gravitating cylindrical structures and found that filaments are unstable to longitudinal perturbations, which grow and cause the filament to fragment into a number of prestellar cores. 

Once a prestellar core has formed, the next step on the way to creating a star is for this core to collapse. Under isothermal conditions, the only requirement for such a collapse is that the core's mass should be larger than its thermal Jeans mass. In reality, however, after sufficient compression the system ceases to be isothermal and instead behaves closer to adiabatic. We must therefore consider the importance of thermodynamics: Is the increase in pressure strong enough to halt the gravitational collapse? 

The following description of protostellar collapse is considerably simplified since it assumes spherical symmetry and ignores magnetic fields and rotation. We will continue to make these assumptions throughout this work, since we focus on 1-D radiation-hydrodynamical simulations. There exist much more sophisticated simulations of protostellar collapse, including for instance magnetic field effects (\citet{Tomisaka2002, Machida2006, Commercon2010,Federrath2015}), protoplanetary disk formation (\citet{inu2010,Stamatellos2008,Machida2011,Tomida2015,Seifried2016,Gonzalez2015,Nordlund2014}), chemical evolution (\citet{Visser2015,Dzyurkevich2016,Hincelin2016}), and non-ideal MHD (\citet{Tsukamoto2015a, Tsukamoto2015b,Wurster2016,Masson2016}). In addition, there are synthetic observations which can be compared to actual data (\citet{Commercon2012a,Frimann2016,Seifried2016}). However, in this study we aim to analyze the effects of initial conditions (mass, radius etc.) on the collapse process in a simplified environment. Our work is rooted in \citet{inu98} and \citet{inu2000}  and it is in many ways similar to \citet{Vaytet2017}, which we will refer to often in this paper. There are two main differences between our work and theirs: First, we focus on a different mass range, namely brown dwarfs and sub-brown dwarfs. This is a mass regime which has so far been underexplored, and we wish to investigate if the results for higher masses (\citet{Vaytet2017}'s lowest mass is $0.2 M_\odot$) also hold for brown dwarfs, or whether there are any new effects. The second difference lies in the radiative transfer scheme. While \citet{Vaytet2017} employ flux-limited radiative diffusion as well as the gray approximation, we use a frequency-dependent method which we developed in our previous paper (\citet{Stamer2018}) specifically for spherically symmetric systems.

In the following, we will give a brief overview of the process of protostellar collapse. We model the cloud core as a homogeneous sphere with total mass $M$, radius $R$ and density $\rho$. We express the pressure using a polytropic equation of state, i.e. $P \propto \rho^{\ix{\gamma}{eff}}$, where $\ix{\gamma}{eff}$ is the effective ratio of specific heats. The gravitational and pressure forces are then given by:

\begin{align}
&\ix{F}{G}=\frac{GM}{R^2} \\
&\ix{F}{P}=\frac{1}{\rho} \frac{\partial P}{\partial R} \propto \frac{\rho^{\ix{\gamma}{eff}-1}}{R}
\end{align}

For a homologous collapse, $\rho \propto R^{-3}$, so the relationship between pressure and radius is $\ix{F}{P} \propto R^{2-3 \ix{\gamma}{eff}}$. The ratio between the two forces then scales as:

\begin{equation}
\frac{\ix{F}{P}}{\ix{F}{G}} \propto R^{4-3\ix{\gamma}{eff}}
\end{equation}

It follows that there is a critical value of $\ix{\gamma}{eff}$, namely 4/3. If $\ix{\gamma}{eff}$ is larger than this value, pressure becomes stronger relative to gravity as the radius decreases, which inevitably leads to a stabilization and stop of the collapse. On the other hand, if it is smaller than this value, gravity becomes ever more dominant over pressure as the radius decreases, leading to an accelerating collapse.

\begin{figure}[htbp]
	\centering
		\plotone{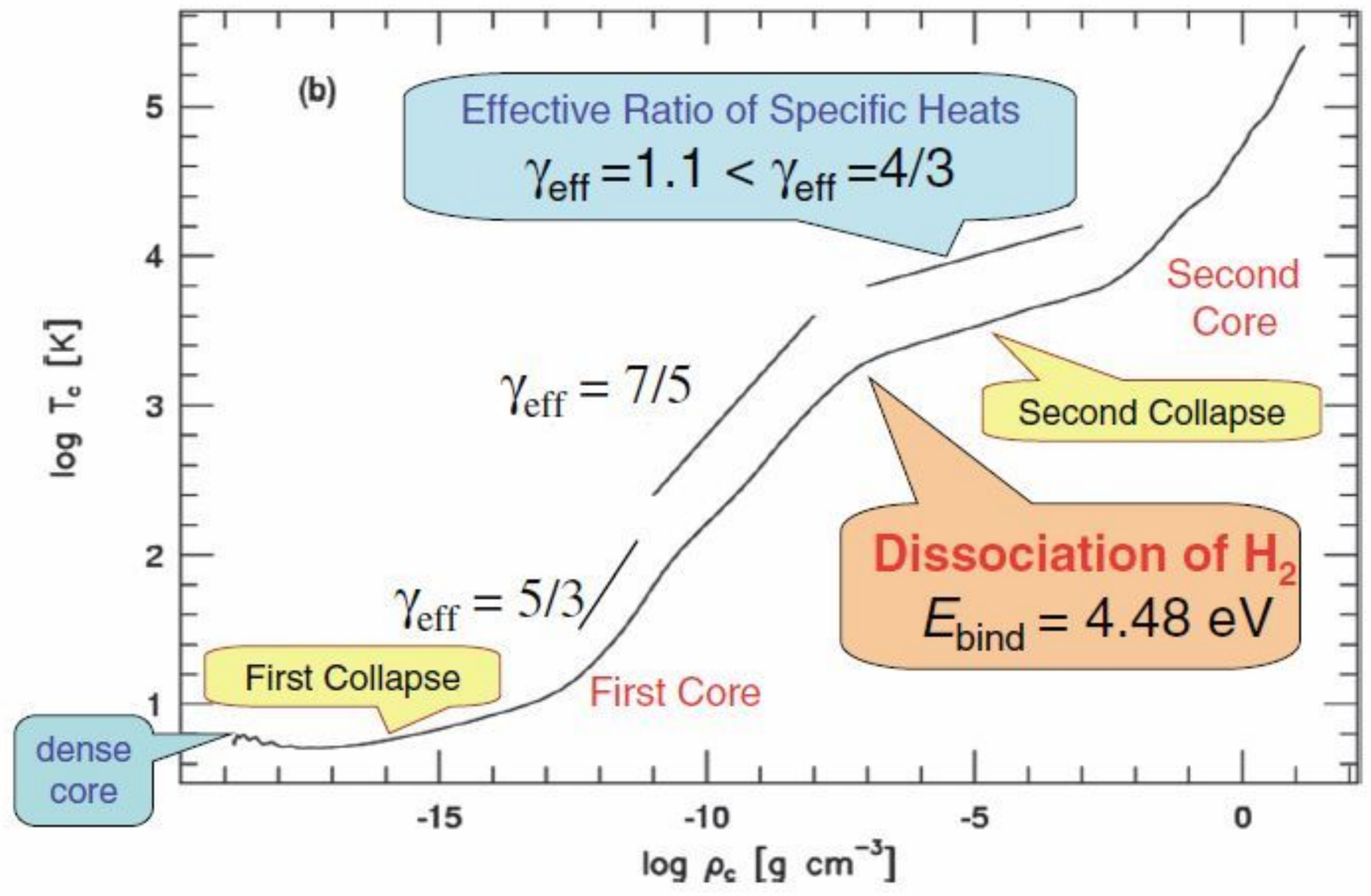}
	\caption{Evolution of temperature and density at the center of a collapsing cloud core, from \citet{inu2012}. An initial isothermal phase is followed by adiabatic contraction until the formation of the first hydrostatic core. The dissociation of hydrogen molecules triggers the second collapse, which results in the formation of the second core (protostar).}
	\label{fig:Collapse}
\end{figure}

We describe the evolution (summarized in Figure \ref{fig:Collapse}): Initially, the cloud core is essentially isothermal ($\ix{\gamma}{eff}=1$), since it is optically thin so that any heat gained through compression is efficiently radiated away and the gas is kept at the temperature of the surrounding molecular cloud, which is typically around 10 K. The temperature actually drops slightly since the core becomes opaque to the high-frequency heating radiation of background stars while the low-frequency cooling radiation is still able to escape, an effect which we will discuss further in sections \ref{sec:CoreEvo} and \ref{sec:nofc}.

As the collapse progresses, compressional heating becomes ever stronger and the efficiency of the radiative cooling diminishes. Once the heating overtakes the cooling, the system's behavior transitions from isothermal to adiabatic. $\ix{\gamma}{eff}$ then becomes 5/3 (for temperatures below a few hundred Kelvin) and 7/5 above that. The difference is due to the fact that for low temperatures, the rotational degree of freedom of the hydrogen molecule is ``frozen out'', preventing it from contributing to the heat capacity. In any case, both of these values are larger than the critical value of 4/3, which means that the pressure force ultimately stops the gravitational contraction and a hydrostatic object with a typical size of a few AU, the so-called ``first hydrostatic core'', is formed. Its surface is marked by a shock in which the infalling material is decelerated. At this point, roughly one free-fall time has passed, which for the densities typical of a solar mass cloud core corresponds to about $\mathrm{10^5}$ years. The first core then continues to increase in mass and temperature as it accretes material from the surrounding gas envelope. Once the temperature reaches $\approx \mathrm{10^3}$ K, hydrogen molecules begin to dissociate. This is a strong cooling effect, which acts to reduce the effective ratio of specific heats to about 1.1 and thereby triggers a second collapse. Because of this, the first core is generally thought to exist for no more than a few thousand years, although theoretical considerations (\citet{Tomida2010}) and the surprisingly large number of candidate objects (e.g. \citet{Pineda2011}, \citet{Dunham2011}, \citet{Chen2010a} and \citet{Chen2010b}) suggest that it may live much longer in some cases. We will discuss this question in detail in section \ref{sec:fclife}.
The second collapse continues until all the hydrogen has been dissociated, so that $\ix{\gamma}{eff}$ increases again to 5/3 and a second hydrostatic core forms, which is also known as a protostar. The collapse phase ends here and the system's pre-main sequence evolution begins.

This paper aims to explore the process of protostellar collapse for very low mass objects, i.e. the brown dwarf and sub-brown dwarf regime. We wish to investigate which of the results that have previously been found for higher-mass objects also hold in this mass range. In addition, we will attempt to interpret the differences that arise as well as their possible impact on observations.  

This paper is organized as follows: In section \ref{sec:Method}, we describe the numerical method used, including the opacities and equation of state. The results of our simulations are then shown and interpreted in section \ref{sec:Results} and further discussed in section \ref{sec:Discussion}, before we summarize our work in section \ref{sec:Conclusion}.

Finally, Appendix \ref{app:GravProb} deals with a numerical problem related to energy conservation in a self-gravitating system. This is an issue that may be important to other researchers doing similar work, and we are unaware of it being discussed in the literature previously, so we analyze it in some detail.

\section{Method}
\label{sec:Method}
\subsection{Basic Equations}
Our hydrodynamics module is based on \citet{Colella1984}), although we use a spatially second-order (piecewise linear) Godunov scheme rather than their piecewise parabolic one. The governing equations of the system in Lagrangian conservation form are as follows: 

\begin{align}
&\frac{dV}{dt}=\frac{d(r^2v)}{dm} \\
&\frac{dv}{dt}=-r^2 \frac{dp}{dm} - \nabla \Phi \\
&\frac{dU}{dt}=-\frac{d(r^2vp)}{dm} - v \nabla \Phi + \left( \frac{dU}{dt} \right)_{\rm rad} \label{eq:E} \\
&\Delta \Phi=\frac{4 \pi G} {V} 
\end{align}

The equations are formulated in terms of the mass coordinate $m(r)$, which is equal to the enclosed mass divided by 4$\pi$. $V$, $v$, $U$, and $\Phi$ are the specific volume, velocity, total specific energy, and gravitational potential. The effect of radiation is included as the term $\left( \frac{dU}{dt} \right)_{\rm rad}$ in the energy equation. This term is calculated according to the radiative transfer scheme for spherically symmetric systems that we developed in \citet{Stamer2018}. This scheme is frequency-dependent (we use 40 frequency bins) and it is very accurate in all regimes of optical depth since it avoids any kind of radiative diffusion approximation. It calculates the radiative quantities in the laboratory frame, but under the assumption that the source function is approximately the same as in the comoving frame. We ignore radiation pressure throughout this work, since it only becomes relevant for very high mass stars, which we are not dealing with here.

\subsection{Equation of State}

\begin{figure}
\plotone{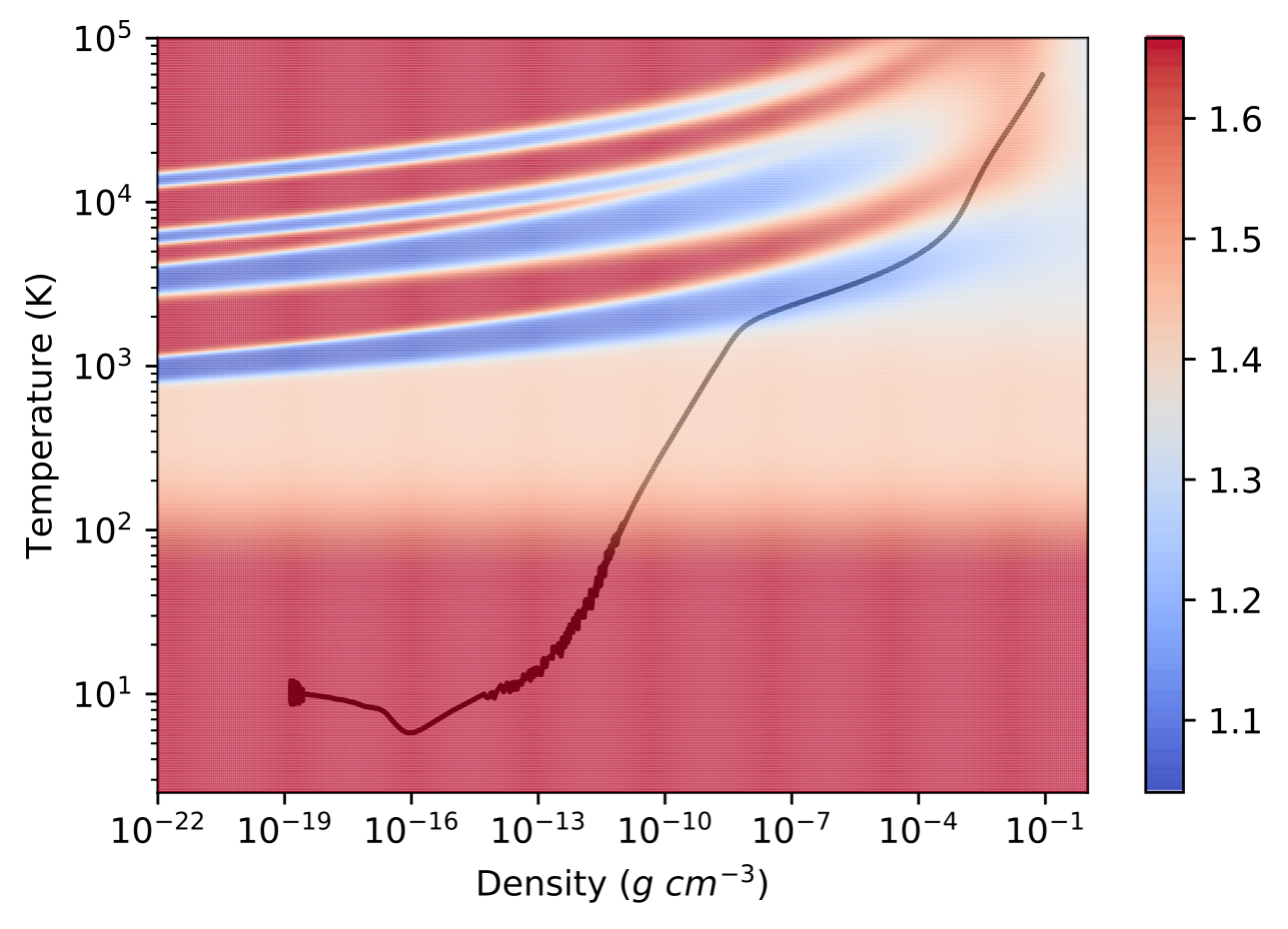}
\caption{Visual representation of the realistic EOS by \citet{Tomida2012}. The black line shows a typical evolutionary track (Run C0) for the central gas element in a protostellar collapse process in the $\rho$-T-plane (see also Figure \ref{fig:Collapse}), while the colored background shows the value of the adiabatic index $\gamma$ at that point. Note the initial decrease in temperature caused by the attenuation of background stellar radiation, which is only captured in frequency-dependent calculations (section \ref{sec:CoreEvo}). The region below about 100K corresponds to $\gamma=5/3$, the value for a monatomic gas, due to the hydrogen molecule's rotational degree of freedom being inactive at these temperatures. At higher temperatures, it decreases to $\gamma=7/5$, the value expected for a diatomic gas. The four bands of lowered $\gamma$ correspond, from bottom to top, to four endothermic reactions: The dissociation of $\ix{H}{2}$, the ionization of hydrogen, and the first and second ionization of helium.}
\label{fig:EOS}
\end{figure}

We use an EOS table developed by Kengo Tomida (Princeton University) and Yasunori Hori (National Astronomical Observatory of Japan), which is described in detail in \citet{Tomida2012} and \citet{Tomida2015} and whose effect is visualized in Figure \ref{fig:EOS}. This table was calculated under the assumptions of local thermodynamical and chemical equilibrium. The latter is only valid if the timescale for chemical reactions is much shorter than the dynamical timescale, but since we are dealing with very dense gas this condition holds. Furthermore, the EOS table assumes solar abundance ($X=0.7$, $Y=0.3$, heavy metals are not considered), and a constant ratio of ortho- to parahydrogen of 3:1. This ratio affects the thermodynamic properties of the gas, but its actual value in molecular cloud cores is poorly constrained. We do not investigate its effect at this point. Finally, as the authors themselves note in the appendix of \citet{Tomida2012}, the EOS becomes unphysical in very high density regions ($\rho>0.1$ g $\mathrm{cm^{-3}}$) since the neglected non-ideal effects are important in that region. This is not an issue for us since our calculations only reach densities about an order of magnitude lower. 

\subsection{Opacity}
We use the monochromatic opacities by \citet{Vaytet2017}, available for download at \url{http://starformation.hpc.ku.dk/?q=grid-of-protostars}. These are a combination of the \citet{Semenov2003} dust opacities for temperatures below 1500 K, molecular opacities calculated based on \citet{Ferguson2005} for the temperature range between 1500 and 3200 K, and atomic opacities from the Opacity Project (\citet{Badnell2005}) above 3200 K.
The number of frequency points is different in each of these three sources of opacity data. For our program, we sample each of them into 40 log-equally spaced bins between $10^{10}$ and $10^{18}$ Hz. The actual opacity data used for a given shell during program execution depends on the shell's density and temperature. For instance, if the temperature is low, the \citet{Semenov2003} data is used, but if it is very high the \citet{Badnell2005} data is used instead.

\subsection{Initial and Boundary Conditions}
The simplest setup for our initial conditions is to envision the cloud core as a self-gravitating, homogeneous, isothermal sphere. In this case, we simply determine an initial density and radius, which together determine the mass. Due to the constant density, this is an extremely unstable setup. We will also consider the case of a sphere that is initially in hydrostatic equilibrium, i.e. a Bonnor-Ebert sphere (\citet{Bonnor1956,Ebert1955}). The condition of hydrostatic equilibrium can be expressed as the following equation:

\begin{equation}
\frac{dP}{dr} = -G \frac{M(r)}{r^2},
\end{equation}

where $M(r)$ is the enclosed mass at radius r. We assume a constant temperature of 10 K and set up the system by fixing the central density and initial radius and numerically integrating this equation outward from the center until that radius. The total mass is then determined, as is $\ix{P}{O}$, the pressure at the outer boundary. An important derived parameter is the Bonnor-Ebert mass $\ix{M}{BE}=1.18 \ix{c}{s}^4 \ix{P}{O}^{-1/2} G^{-3/2}$, which is the critical mass above which the core is unstable to gravitational collapse. Thus for all our calculations, $M>\ix{M}{BE}$. 

The boundary condition at the center is $v=0$, while the outer boundary is infalling. This infall velocity is calculated from the velocity of the outermost shell, under the assumption that $v$ in that shell varies linearly in space at a given timestep. For the radiative boundary condition, we employ the setup of \citet{inu2000}, where the heating from the outside is a combination of three separate contributions:

\begin{itemize}
\item Heating due to cosmic rays, for which we adopt the value of $6.4 \times 10^{-28} n(\ix{H}{2})$ erg $\mathrm{m^{-3}}$ $\mathrm{s^{-1}}$ (\citet{Goldsmith1978}). This is implemented simply as an additional, constant term in the energy equation. For this purpose, we assume that the gas consists entirely of $H_2$.
\item A 2.7 K blackbody radiation representing the cosmic microwave background.
\item A 6000 K blackbody radiation representing background stellar photons, multiplied by a dilution factor $(10/6000)^x$. We set $x=4.9$, which together with the other two heating sources keeps the initial cloud close to 10K.
\end{itemize}

Following \citet{Vaytet2017}, we stop each run once the density in the central shell decreases from one time step to the next for the first time after the beginning of the second collapse. There are several reasons for this: First, it allows a comparison of different runs at the same point during their evolution, namely the point where the second collapse ends and a first ``hydrostatic bounce'' occurs. Second, our EOS becomes invalid at densities above $\approx 0.1$ g $\mathrm{cm^{-3}}$, so we should stop the calculation before such densities are reached. And finally, the main accretion phase that follows the second collapse is difficult to deal with in a Lagrangian scheme because the time steps become extremely short, on the order of hours and less, and the shells inside the second core become extremely thin, even though we do not really need a high spatial resolution there. For this reason, an effective treatment of the further evolution requires a change in method, such as switching to a Eulerian grid or using an implicit scheme. Finally, it should be noted that the exact density at which the hydrostatic bounce occurs depends on the resolution.

\begin{figure}
\plotone{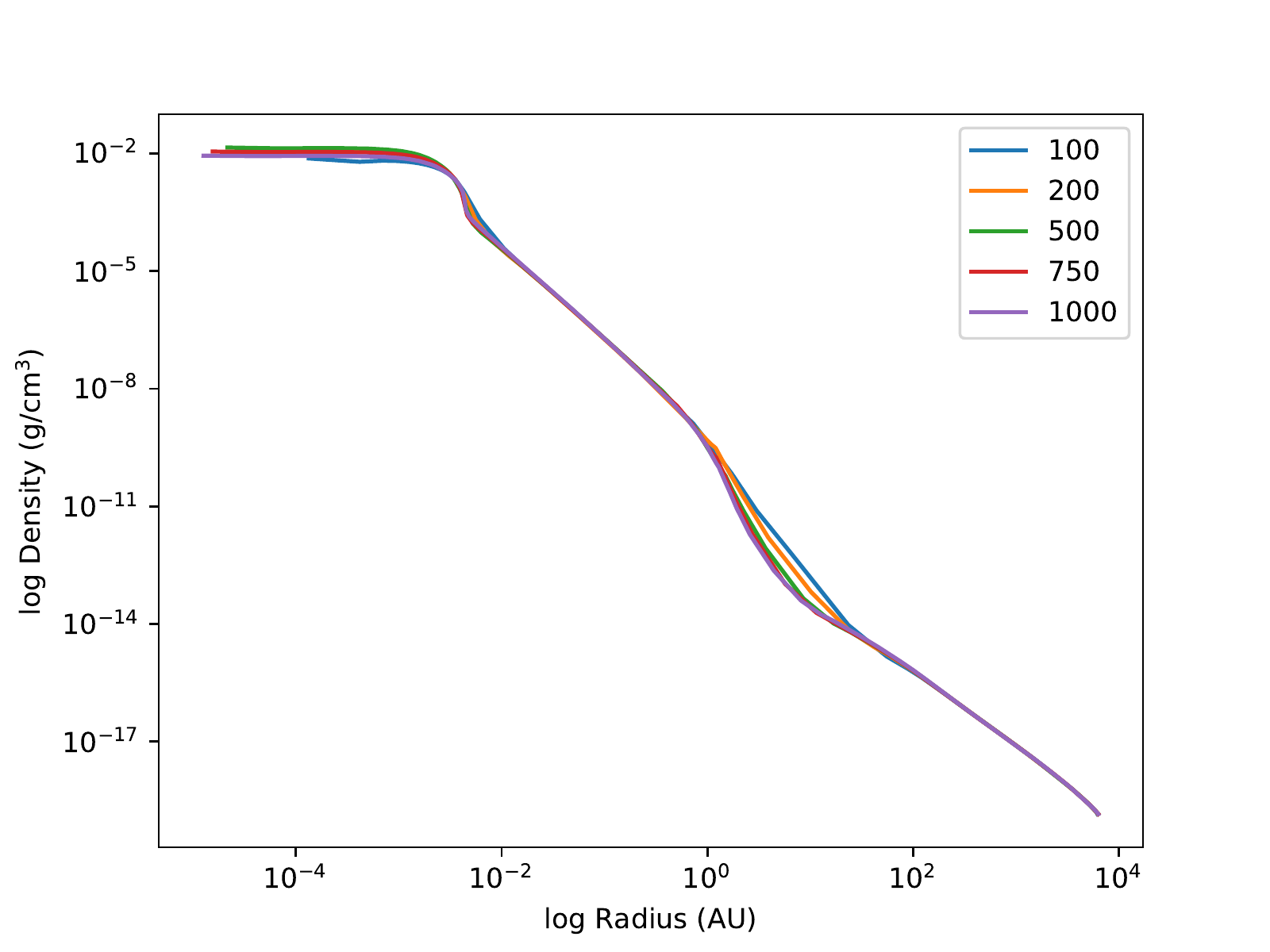}
\caption{Final density profiles for run C0, using different resolutions. The legend indicates the number of shells used.}
\label{fig:resolution}
\end{figure}

We perform simulation runs with initial central densities between $5 \times 10^{-17}$ and $5 \times 10^{-15}$ g $\mathrm{cm^{-3}}$ and initial radii between 65 and 750 AU. These are very dense and compact objects, which is necessary for them to be gravitationally unstable. The masses are between 0.8 and 10 percent of the solar mass, covering the brown dwarf and some of the sub-brown dwarf mass regime (the deuterium burning limit, conventionally used as the lower bound for brown dwarfs, lies at $\approx$ 1.3 percent of solar mass). 

We consider two different initial density structures: One is a simple homogeneous density, which makes for a very unstable setup. The labels of these runs begin with the letter A. The second setup is a hydrostatic, Bonnor-Ebert sphere profile. The labels of these runs begin with the letter B. In addition, we include a one solar mass run with initial conditions identical to \citet{inu2000} for comparison (run C0), as well as a number of runs to test for first core-less collapses (runs D0 through D3; see section \ref{sec:nofc}). A complete overview of all the runs is given in Table \ref{tab:Runs}.

\begin{sidewaystable}
\centering
\begin{tabular}{c | c | c | c | c | c | c | c | c | c}
Run & Profile & $\ix{\rho}{C}$ (g $\mathrm{cm^{-3}})$ & $\ix{R}{O} (AU)$ & $M (\ix{M}{\odot})$ & $\frac{M}{\ix{M}{BE}}$ & $\ix{t}{ff} (yr)$ & $\ix{t}{FC} (yr)$ & $\ix{\tau}{FC} (yr)$ & $\ix{\rho}{fin}$ (g $\mathrm{cm^{-3}})$ \\
\hline
A0 & Hom & $5 \times 10^{-15}$ & 58 & $7 \times 10^{-3}$ & 2.4 & $9.42\times 10^2$ & $1.22\times 10^3$ & $1.49\times 10^4$ & $1.44\times 10^{-2}$ \\
A1 & Hom & $1 \times 10^{-15}$ & 104 & $8 \times 10^{-3}$ & 1.24 & $2.11\times 10^3$ & $3.06\times 10^3$ & $6.81\times 10^3$ & $1.29\times 10^{-2}$ \\
A2 & Hom & $1 \times 10^{-15}$ & 112 & $1.0 \times 10^{-2}$ & 1.54 & $2.11\times 10^3$ & $2.67\times 10^3$ & $3.87\times 10^3$ & $8.09\times 10^{-3}$ \\
A3 & Hom & $1 \times 10^{-15}$ & 119 & $1.2 \times 10^{-2}$ & 1.85 & $2.11\times 10^3$ & $2.52\times 10^3$ & $3.96\times 10^3$ & $7.24\times 10^{-3}$ \\
A4 & Hom & $1 \times 10^{-15}$ & 129 & $1.5 \times 10^{-2}$ & 2.36 & $2.11\times 10^3$ & $2.36\times 10^3$ & $2.27\times 10^3$ & $8.17\times 10^{-3}$ \\
A5 & Hom & $5 \times 10^{-16}$ & 181 & $2.1 \times 10^{-2}$ & 2.30 & $2.98\times 10^3$ & $3.25\times 10^3$ & $9.04\times 10^2$ & $4.59\times 10^{-3}$ \\
A6 & Hom & $5 \times 10^{-17}$ & 463 & $3.5 \times 10^{-2}$ & 1.22 & $9.42\times 10^3$ & $1.04\times 10^4$ & $9.91\times 10^2$ & $7.33\times 10^{-3}$ \\
A7 & Hom & $5 \times 10^{-17}$ & 548 & $5.8 \times 10^{-2}$ & 2.02 & $9.42\times 10^3$ & $9.56\times 10^3$ & $6.38\times 10^2$ & $5.83\times 10^{-3}$ \\
A8 & Hom & $5 \times 10^{-17}$ & 586 & $7.1 \times 10^{-2}$ & 2.47 & $9.42\times 10^3$ & $9.44\times 10^3$ & $5.45\times 10^2$ & $5.83\times 10^{-3}$ \\
A9 & Hom & $5 \times 10^{-17}$ & 622 & $8.5 \times 10^{-2}$ & 2.95 & $9.42\times 10^3$ & $9.36\times 10^3$ & $4.59\times 10^2$ & $6.56\times 10^{-3}$ \\
A10 & Hom & $5 \times 10^{-17}$ & 659 & $1.0 \times 10^{-1}$ & 3.51 & $9.42\times 10^3$ & $9.33\times 10^3$ & $3.64\times 10^2$ & $6.58\times 10^{-3}$ \\
B0 & BE & $5 \times 10^{-15}$ & 65 & $7 \times 10^{-3}$ & 1.93 & $1.10\times 10^3$ & $1.54\times 10^3$ & $1.09\times 10^4$ & $1.28\times 10^{-2}$ \\
B1 & BE & $1 \times 10^{-15}$ & 110 & $8 \times 10^{-3}$ & 1.04 & $2.31\times 10^3$ & $3.83\times 10^3$ & $1.45\times 10^4$ & $1.02\times 10^{-2}$ \\
B2 & BE & $1 \times 10^{-15}$ & 120 & $1.0 \times 10^{-2}$ & 1.28 & $2.35\times 10^3$ & $3.31\times 10^3$ & $4.78\times 10^3$ & $9.14\times 10^{-3}$ \\
B3 & BE & $1 \times 10^{-15}$ & 130 & $1.2 \times 10^{-2}$ & 1.53 & $2.39\times 10^3$ & $3.07\times 10^3$ & $2.58\times 10^3$ & $9.16\times 10^{-3}$ \\
B4 & BE & $1 \times 10^{-15}$ & 140 & $1.5 \times 10^{-2}$ & 1.79 & $2.43\times 10^3$ & $2.93\times 10^3$ & $2.07\times 10^3$ & $1.15\times 10^{-2}$ \\
B5 & BE & $5 \times 10^{-16}$ & 200 & $2.1 \times 10^{-2}$ & 1.83 & $3.45\times 10^3$ & $3.92\times 10^3$ & $8.21\times 10^2$ & $6.30\times 10^{-3}$ \\
B6 & BE & $5 \times 10^{-17}$ & 500 & $3.5 \times 10^{-2}$ & 1.08 & $1.04\times 10^4$ & $1.20\times 10^4$ & $1.06\times 10^3$ & $8.97\times 10^{-3}$ \\
B7 & BE & $5 \times 10^{-17}$ & 600 & $5.8 \times 10^{-2}$ & 1.64 & $1.08\times 10^4$ & $1.11\times 10^4$ & $6.95\times 10^2$ & $8.97\times 10^{-3}$ \\
B8 & BE & $5 \times 10^{-17}$ & 650 & $7.1 \times 10^{-2}$ & 1.93 & $1.10\times 10^4$ & $1.10\times 10^4$ & $6.39\times 10^2$ & $6.59\times 10^{-3}$ \\
B9 & BE & $5 \times 10^{-17}$ & 700 & $8.5 \times 10^{-2}$ & 2.24 & $1.12\times 10^4$ & $1.09\times 10^4$ & $5.50\times 10^2$ & $8.23\times 10^{-3}$ \\
B10 & BE & $5 \times 10^{-17}$ & 750 & $1.0 \times 10^{-1}$ & 2.55 & $1.15\times 10^4$ & $1.08\times 10^4$ & $5.67\times 10^2$ & $9.31\times 10^{-3}$ \\
C0 & Hom & $1.42 \times 10^{-19}$ & 10000 & $ 1 $ & 1.85 & $1.77\times 10^5$ & $ 1.85\times 10^5 $ & $ 9.01\times 10^2$ & $1.33\times 10^{-2}$ \\
D0 & Hom & $3.28 \times 10^{-19}$ & 6000 & $ 0.5 $ & 1.41 & $1.16\times 10^5$ & $1.28\times 10^5$ & $8.78 \times 10^2$ & $1.57\times 10^{-2}$ \\
D1 & Hom & $3.28 \times 10^{-19}$ & 8000 & $ 1.18 $ & 3.34 & $1.16\times 10^5$ & $1.28\times 10^5$ & $6.20 \times 10^2$ & $1.13\times 10^{-2}$ \\
D2 & Hom & $3.28 \times 10^{-19}$ & 11000 & $ 3.08 $ & 8.68 & $1.16\times 10^5$ & - & - & $3.77\times 10^{-3}$ \\
D3 & Hom & $3.28 \times 10^{-19}$ & 12000 & $ 4 $ & 11.27 & $1.16\times 10^5$ & - & - & $3.80\times 10^{-3}$ \\
\end{tabular}
\caption{Overview of our simulation runs. The values are, from left to right: Run label, density profile, central density at the beginning of the simulation, initial radius, mass, ratio of mass to Bonnor-Ebert mass, free-fall time, time until first core formation, first core lifetime, central density at the end of the run. The time of first core formation is defined as the first timestep where the density in the central shell decreases from one time step to the next. The first core lifetime is the time elapsed between that same timestep and the end of the simulation.}
\end{sidewaystable}
\label{tab:Runs}

\subsection{Numerical Resolution}
Figure \ref{fig:resolution} shows the final density profiles for our run C0 for different resolutions between 100 and 1000 shells. We find that the resolution has a slight effect on the profile, especially near the first core region. For more than 500 shells, further changes are very small, but require a significant increase in computational time. One of our most significant results concerns the lifetime of the first core (section \ref{sec:fclife}). For very low resolutions, its value increases significantly (1550 years for 50 shells, 1100 years for 100 shells), but converges to about 900 years at 200 shells and more. We use 500 shells for all the calculations discussed in this paper.

\section{Results}
\label{sec:Results}

\subsection{Core Evolution}
\label{sec:CoreEvo}
The evolution in the temperature-density plane of a fluid element near the center of the system was already shown in Figure \ref{fig:EOS}. It generally behaves as expected (see also Figure \ref{fig:Collapse}), with a near-isothermal phase until $\approx 10^{-12}$ g $\mathrm{cm^{-3}}$, followed by an adiabatic phase until  $\approx 10^{-8}$ g $\mathrm{cm^{-3}}$ and the second collapse until $\approx 10^{-3}$ g $\mathrm{cm^{-3}}$, when hydrogen molecule dissociation is complete and the equation of state again becomes stiffer. 

It is noteworthy, however, that the initial contraction is not completely isothermal. The temperature oscillations at densities below $10^{-18}$ g $\mathrm{cm^{-3}}$ are a numerical artifact, but the following temperature decrease is significant. The temperature drops from the initial value of $\approx 10$ K to a minimum of $\approx 6$ K before increasing again. This effect is only captured in frequency-dependent radiative transfer calculations such as this work or \citet{inu2000}, but not in gray cases such as \citet{Vaytet2017} and \citet{Tomida2010}. The cause is due to the increased opacity for radiation from the outside: As the cloud core starts to contract, its optical thickness begins to increase. Since the opacity is generally larger at higher frequencies, the system first becomes optically thick to high-frequency photons while still remaining optically thin to low frequencies.  The inner regions are thus shielded from the background stellar radiation, so that heating is supplied only by cosmic rays and the CMB, but the cooling radiation emitted from the dust can still escape. In conclusion, a numerical simulation must satisfy two conditions to show this effect: It must be frequency-dependent, and it must include high-frequency radiation from the outside (such as stellar background radiation) as a heating source.

\subsection{Radial Profiles}
\begin{figure}
\plotone{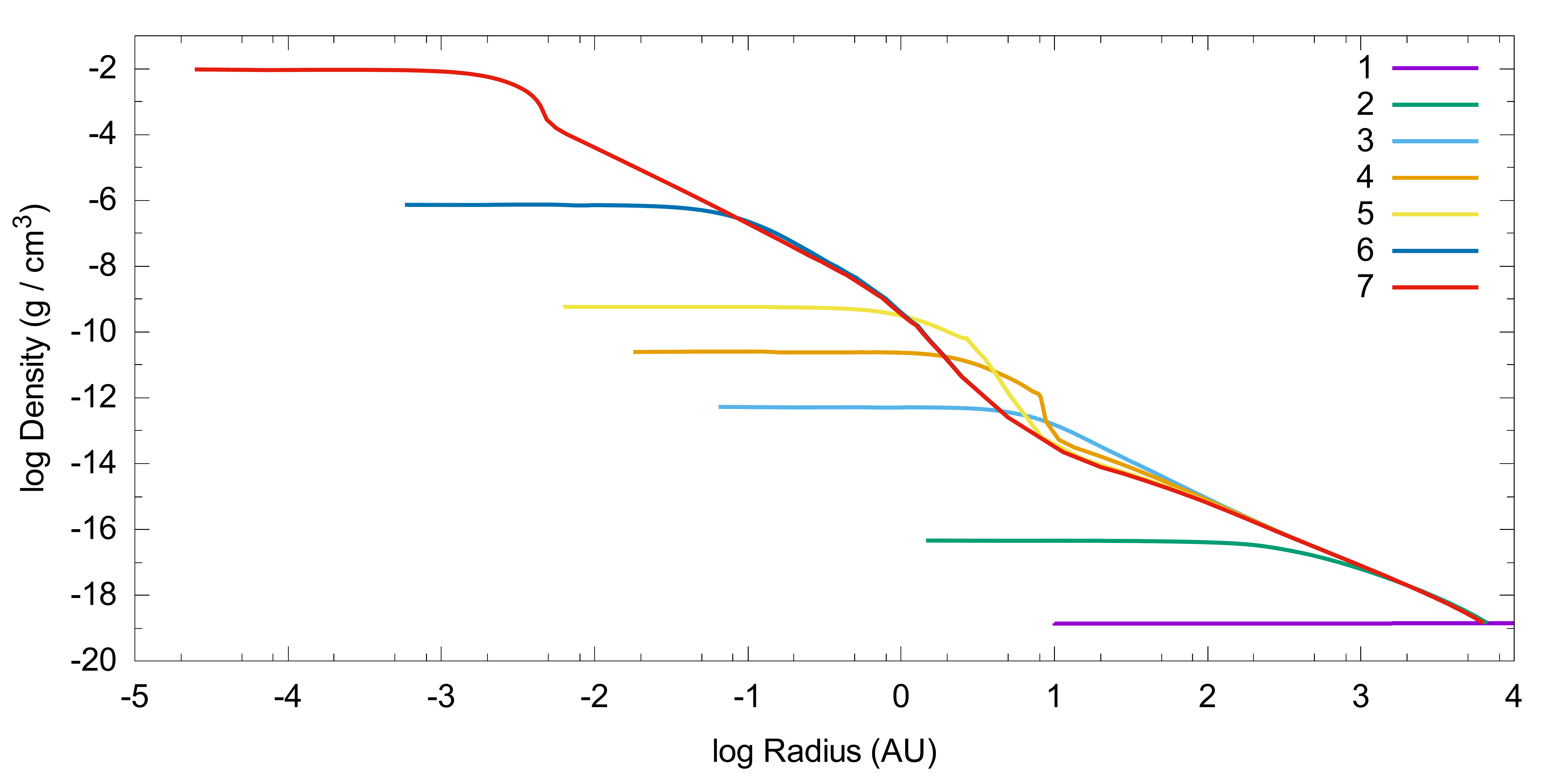}
\plotone{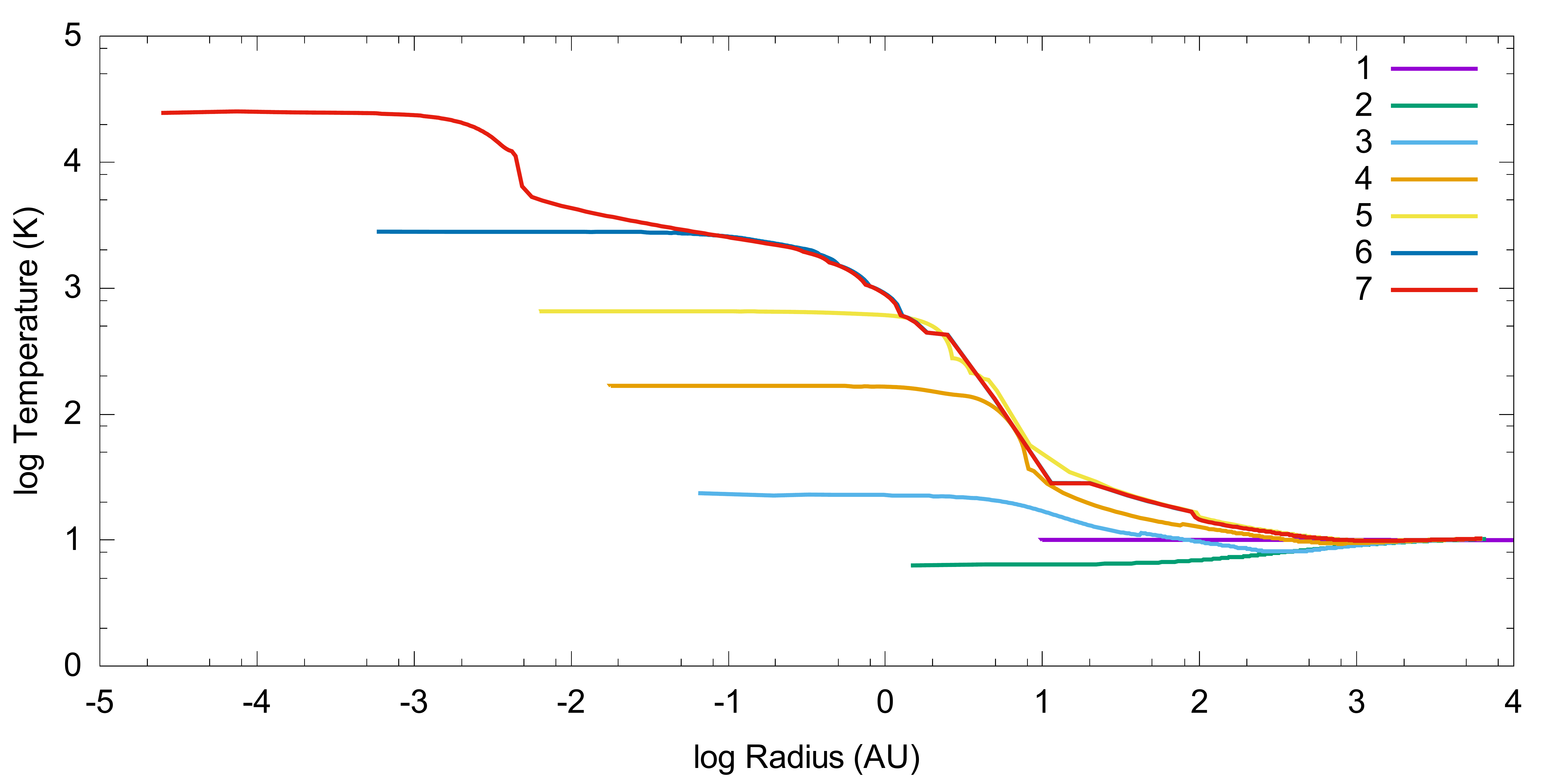}
\plotone{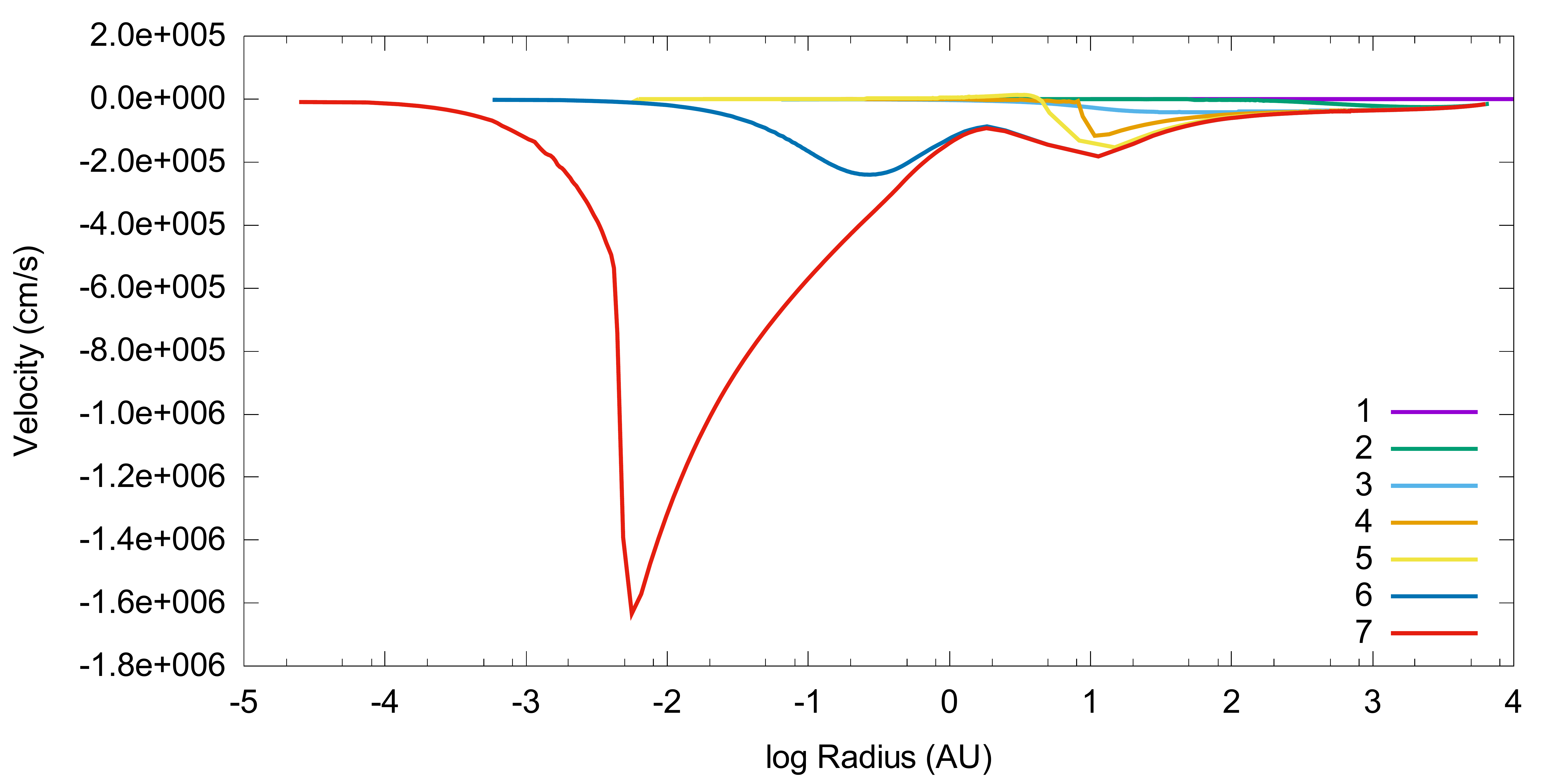}
\caption{Evolution of the radial profiles of density, temperature and velocity for run C0. 7 snapshots are shown, numbered sequentially after their elapsed time. In terms of physical time, it takes about one free-fall time ($\approx$180,000 years) until the formation of the first core (steps 1 to 4), but only about 1000 years from step 4 to 7.}
\label{fig:Radial}
\end{figure}

The evolution of run C0's radial profiles of density, temperature and velocity is plotted in Figure \ref{fig:Radial}. The formation of the first and second cores at $\approx 10^1$ and $\leq 10^{-2}$ AU, respectively,  is evident as a sudden increase in density and temperature and as a maximum in the infall velocity (accretion shock). Snapshot 2 also shows the initial temperature decrease mentioned in the previous section, and it can be seen that this decrease affects the whole inner region of the cloud core, but not the outermost regions.

\begin{figure}
\plottwo{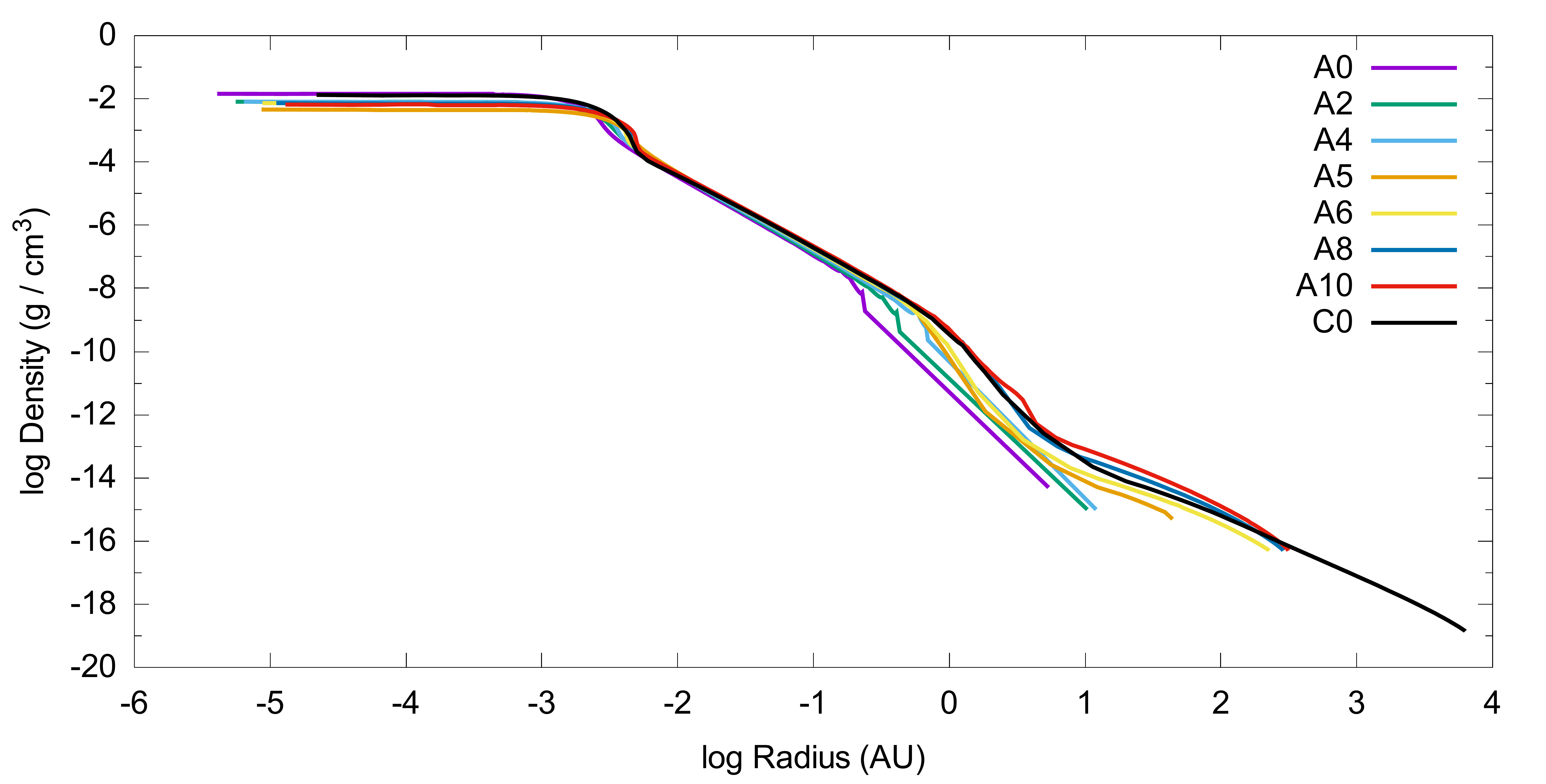}{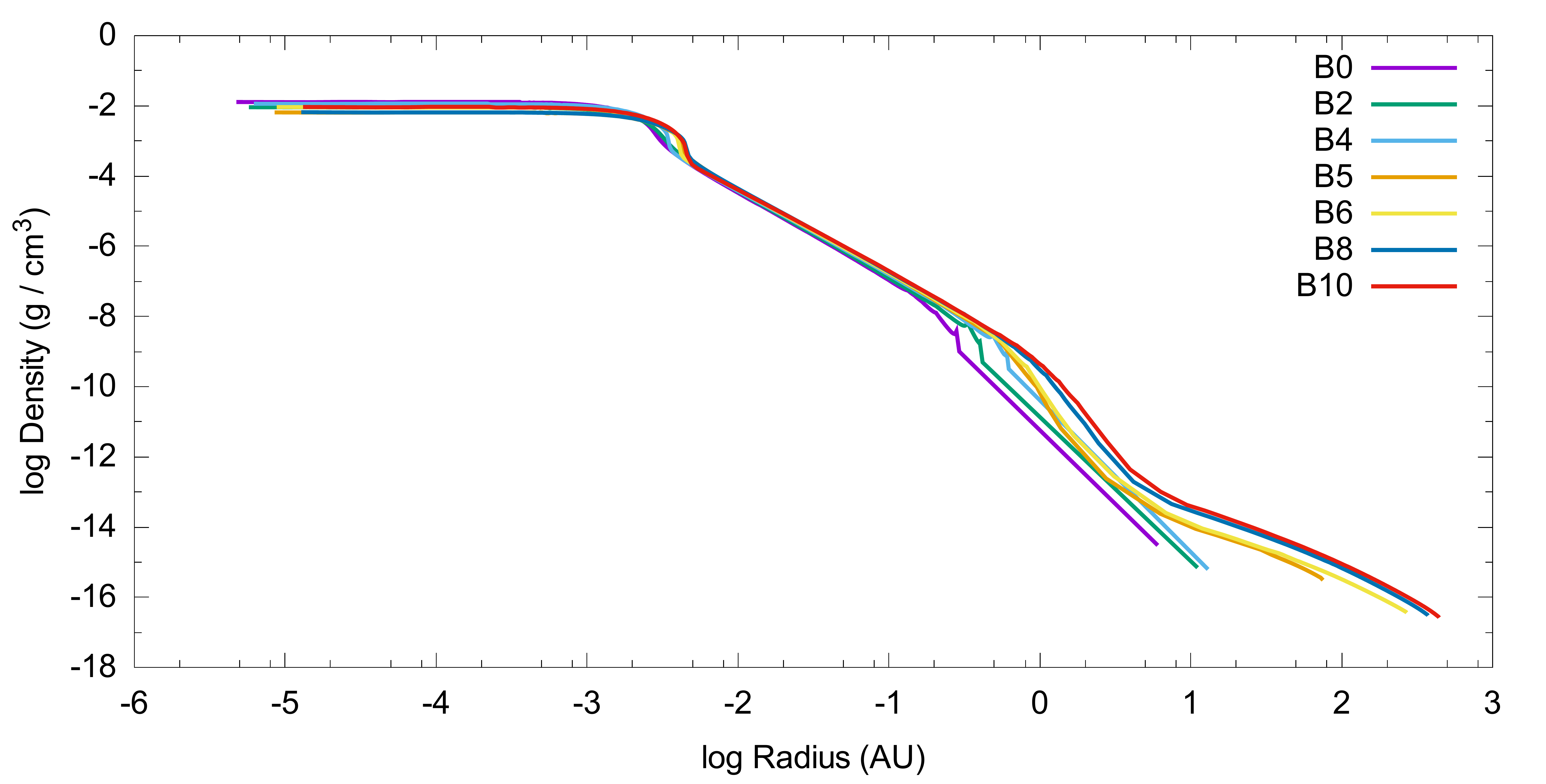}
\plottwo{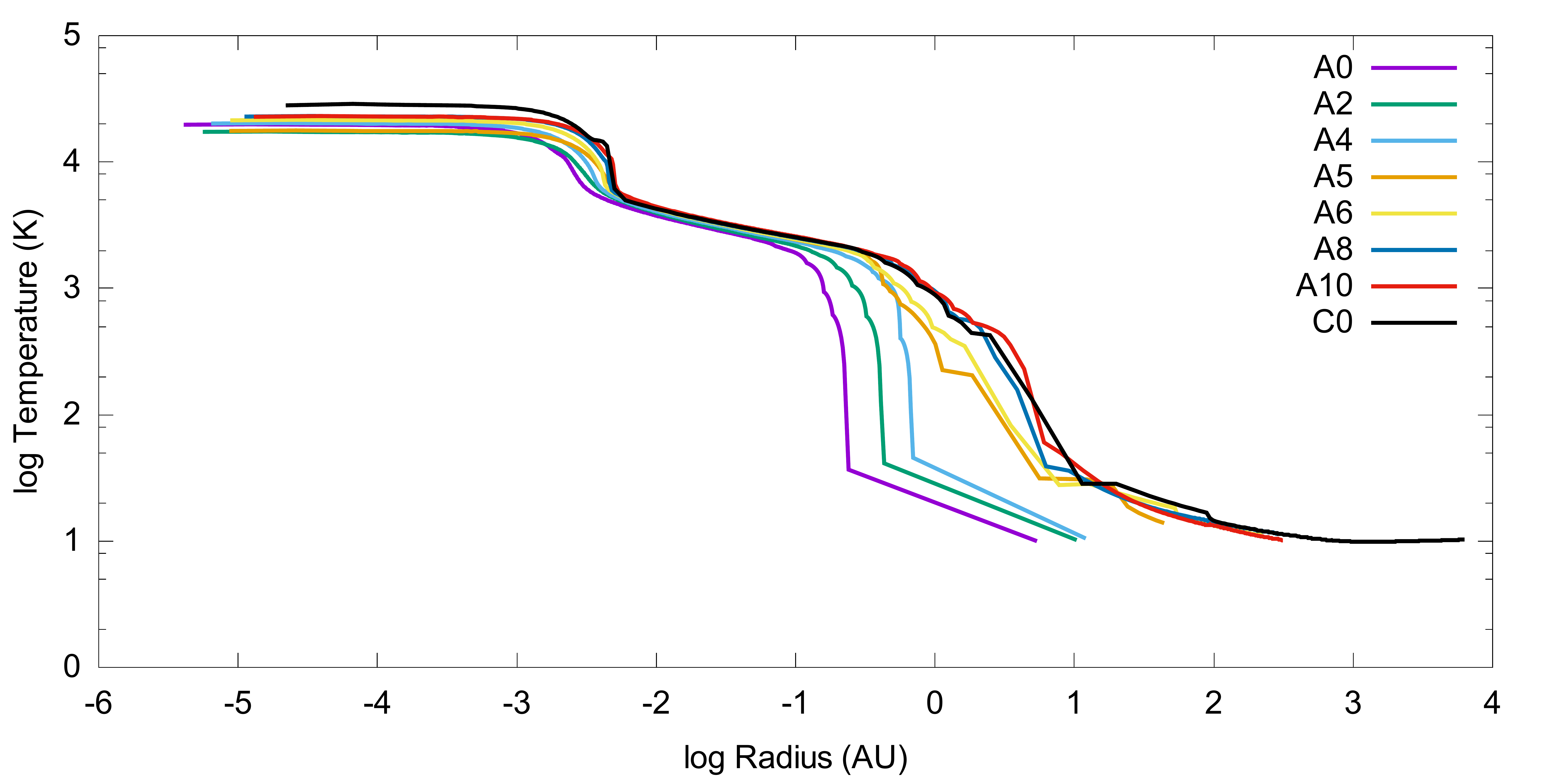}{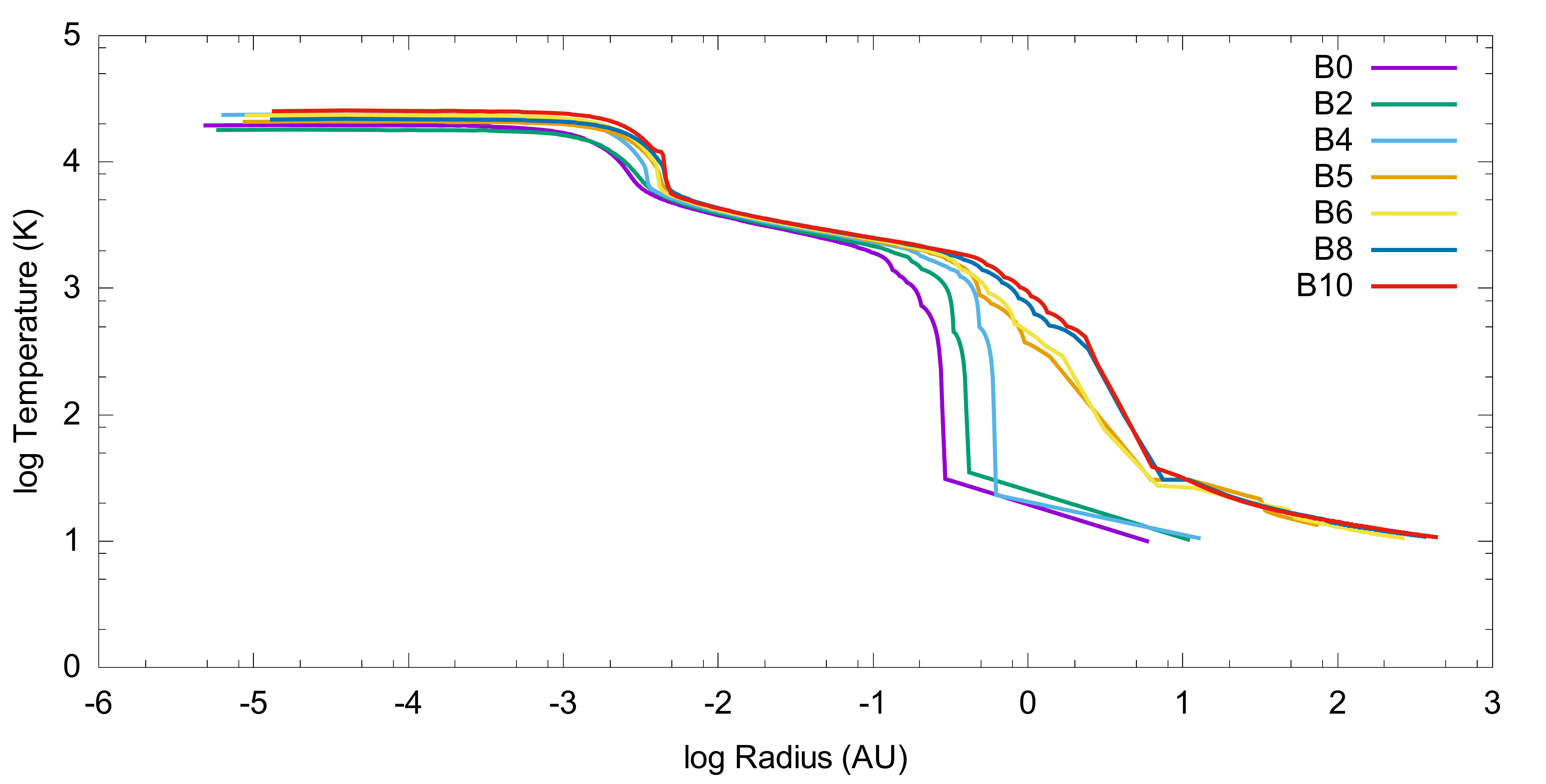}
\plottwo{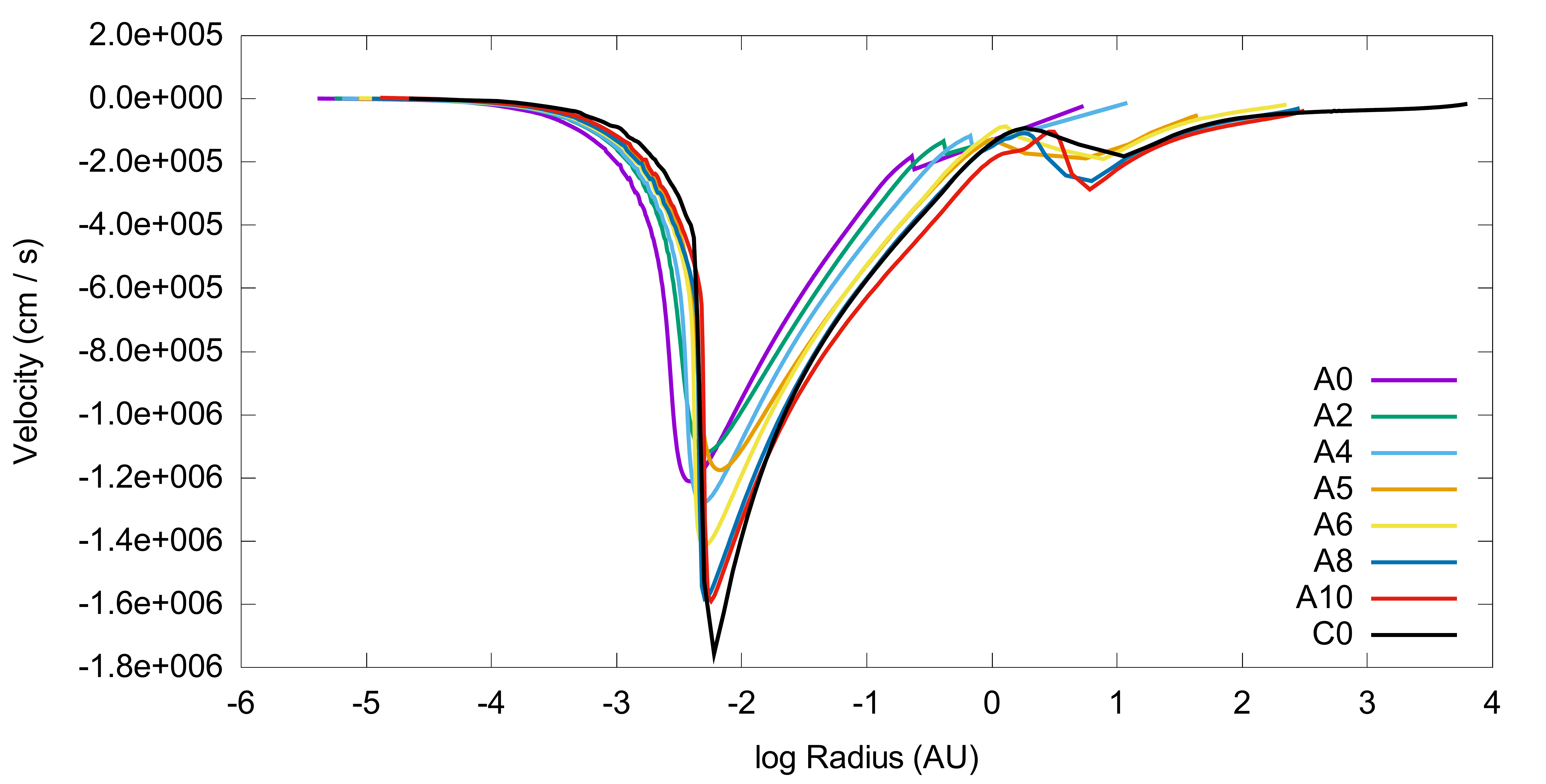}{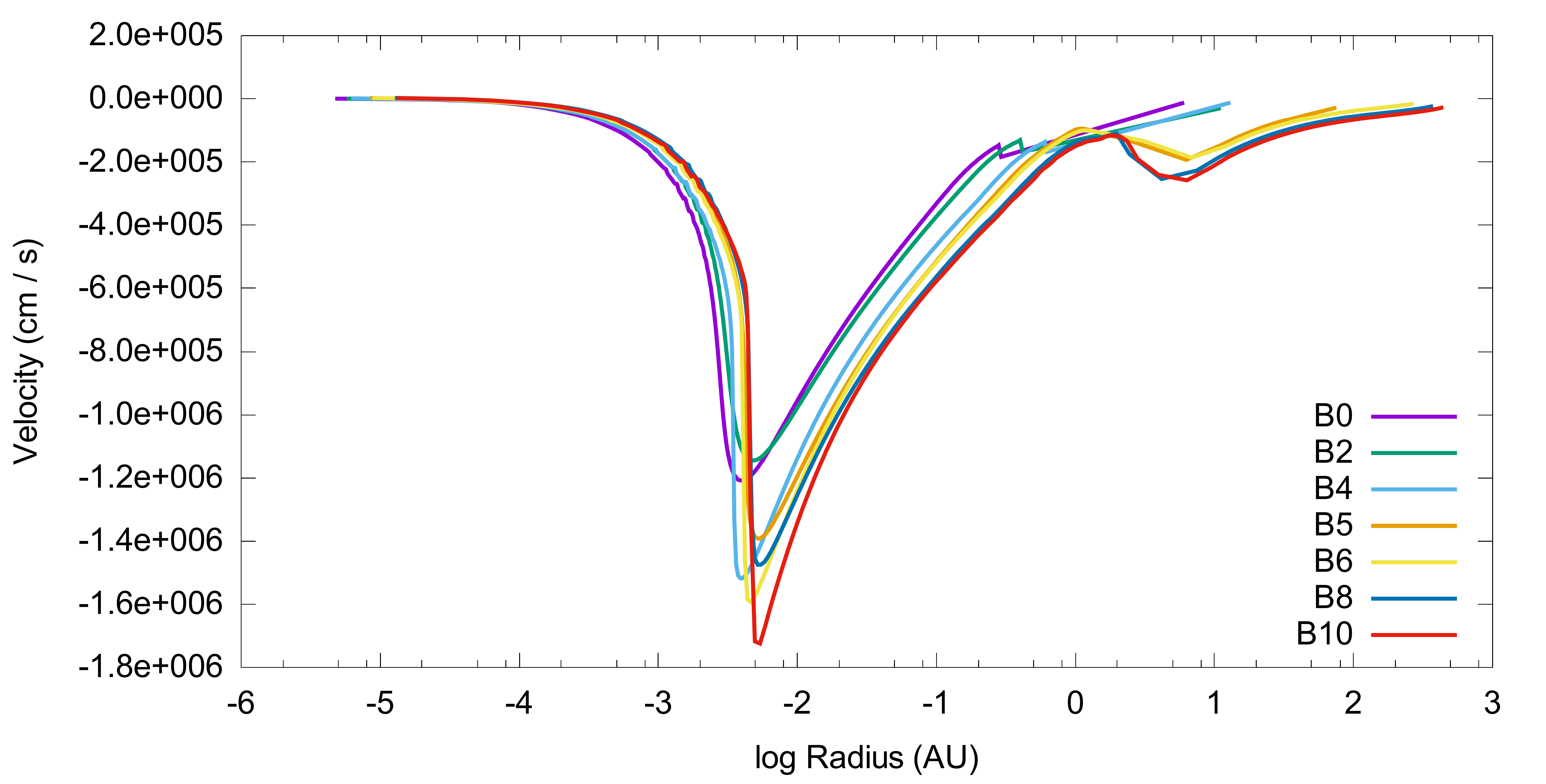}
\caption{Density, temperature and velocity profiles at the end of the simulation for various runs. The left side shows runs with an initially homogeneous density profile, while the right side shows initially Bonnor-Ebert profiles.}
\label{fig:Finals}
\end{figure}

The final radial profiles (Figure \ref{fig:Finals}) show that the second core's properties (density, radius) are essentially independent of the initial conditions. They are all very similar irrespective of the cloud core mass, initial density and density structure. This agrees with the results of \citet{Vaytet2017} for higher masses, but here we show that it holds true in the brown dwarf regime as well.

However, we do find significant differences in the first core. Specifically, the very low mass runs A0 - A4 and B0 - B4 look qualitatively different, with a very small first core ($<$ 1 AU) which essentially makes up the whole system. The unusual behavior of these runs is closely related to the question of first core lifetime and is discussed in the following section.

\subsection{First Core Lifetimes}
\label{sec:fclife}
The first hydrostatic core is generally believed to be a very short-lived object, which combined with the low luminosity should make it very hard to observe. Nevertheless, as was mentioned in the introduction, a relatively large number of first core candidates have been observed. While none of these are confirmed to be first cores, and it is entirely possible that some or all of them may turn out to be more evolved objects, this large number of candidate observations may also indicate that at least some first cores are more long-lived and therefore more common than typically believed. \citet{Tomida2010} investigated first core lifetimes using 3D RHD simulations and found that for low mass cores ($<0.1 \ix{M}{\odot}$), the lifetime can be in excess of $10^4$ years. The reason for this is that for the second collapse to start, the first core must reach a critical temperature of about 2000 K. How quickly that temperature is reached depends on the mass accretion rate. However, in a very low-mass case the mass reservoir may essentially be used up so that the accretion onto the first core becomes very weak before the temperature has risen sufficiently. This is what we see in the radial profiles of the very low mass runs A0 - A4 and B0 - B4 (Figure \ref{fig:Finals}): The first core basically makes up the whole system. If - somewhat arbitrarily - we define the surrounding envelope as the region where the temperature is below 100 K, then for example in the final profile of run B10, 44\% of the total mass remain in the envelope. For run B0, it is a mere 2\%. In this situation, radiative cooling then plays a crucial role in removing entropy from the system and thus allowing further contraction and heating in a Kelvin-Helmholtz-like process on a much slower timescale. 

As shown in table \ref{tab:Runs}, our calculations confirm that the first core lifetime rises drastically for these very low-mass runs, when it becomes comparable to or even much larger than the free-fall time. However, this only becomes evident below about $0.02 \ix{M}{\odot}$, not $0.1 \ix{M}{\odot}$ as suggested by \citet{Tomida2010}, and a lifetime of $10^4$ years is only reached in the sub-brown dwarf regime. This is likely to be due to differences in the numerical method: As a 3-D calculation, \citet{Tomida2010} include the effect of rotation, but on the other hand they use an idealized EOS with $\gamma=5/3$ and a much more simplified radiative transfer scheme (gray flux-limited diffusion approximation). 

\begin{figure}
\plotone{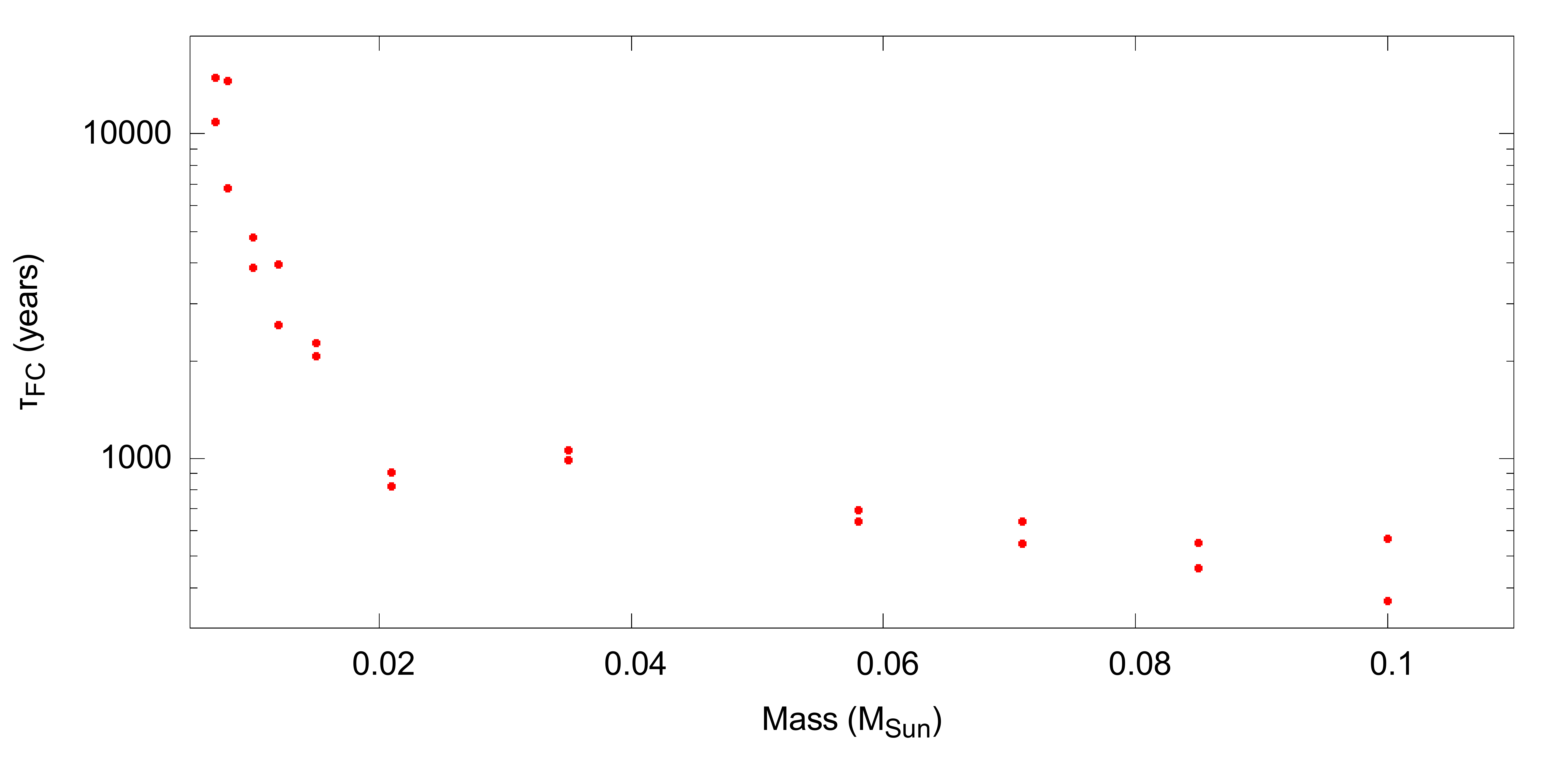}
\plotone{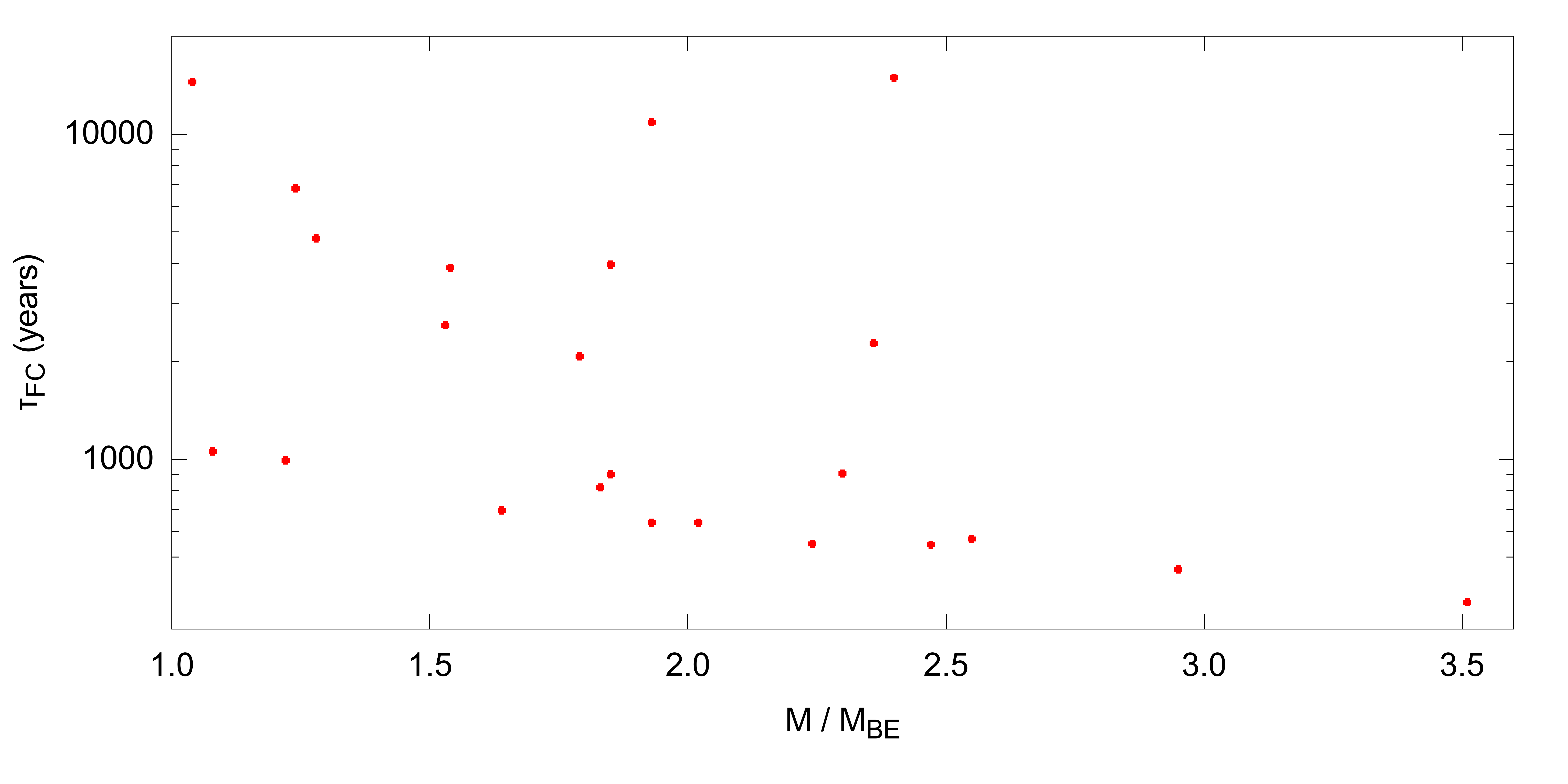}
\caption{\textit{Upper Panel}: Correlation of the first core lifetime $\ix{\tau}{FC}$ with the mass. While there is little effect at masses larger than about $10^{-2} M_{\odot}$, a very significant increase in the lifetime can be observed at extremely low masses. \textit{Lower Panel}: Correlation of $\ix{\tau}{FC}$ with the degree of instability. A decreasing tendency with increasing instability is still discernible, but the correlation is much less pronounced than for the mass.}
\label{fig:FCLifetime}
\end{figure}

\citet{Vaytet2017} find an anticorrelation of the first core lifetime with an instability parameter given by the ratio of mass to Bonnor-Ebert mass, where the lifetime drops at first slowly and then significantly for very large ratios of $M / \ix{M}{BE} > \approx 5$. However, the lowest mass they consider is 0.2 $\ix{M}{\odot}$, so they do not probe the very low mass regime, where the behavior is different. Figure \ref{fig:FCLifetime} shows that for the very low-mass end, the first core lifetime depends strongly on the mass of the cloud core rather than on the instability parameter.

In conclusion, despite the quantitative differences with \citet{Tomida2010}, our results confirm their prediction that very low-mass first cores can be very long-lived due to exhaustion of the accretion reservoir, and that generally the behavior of the first core lifetime is qualitatively different from higher mass ranges. This may account for the surprisingly large number of observational candidates. Furthermore, we predict that future discoveries of first cores will show a bias towards low mass objects due to their longer lifetimes.

\subsection{Collapses Without First Cores}
\label{sec:nofc}

\citet{Vaytet2017} found that for certain very unstable setups ($M / \ix{M}{BE} > \approx 10$), the collapse can proceed without the formation of a first core. This occurs because the vigorous collapse creates such a high ram pressure that the thermal pressure does not become strong enough to balance it before the temperature reaches 2000 K and thus directly proceeds to the second collapse.

\begin{figure}
\plotone{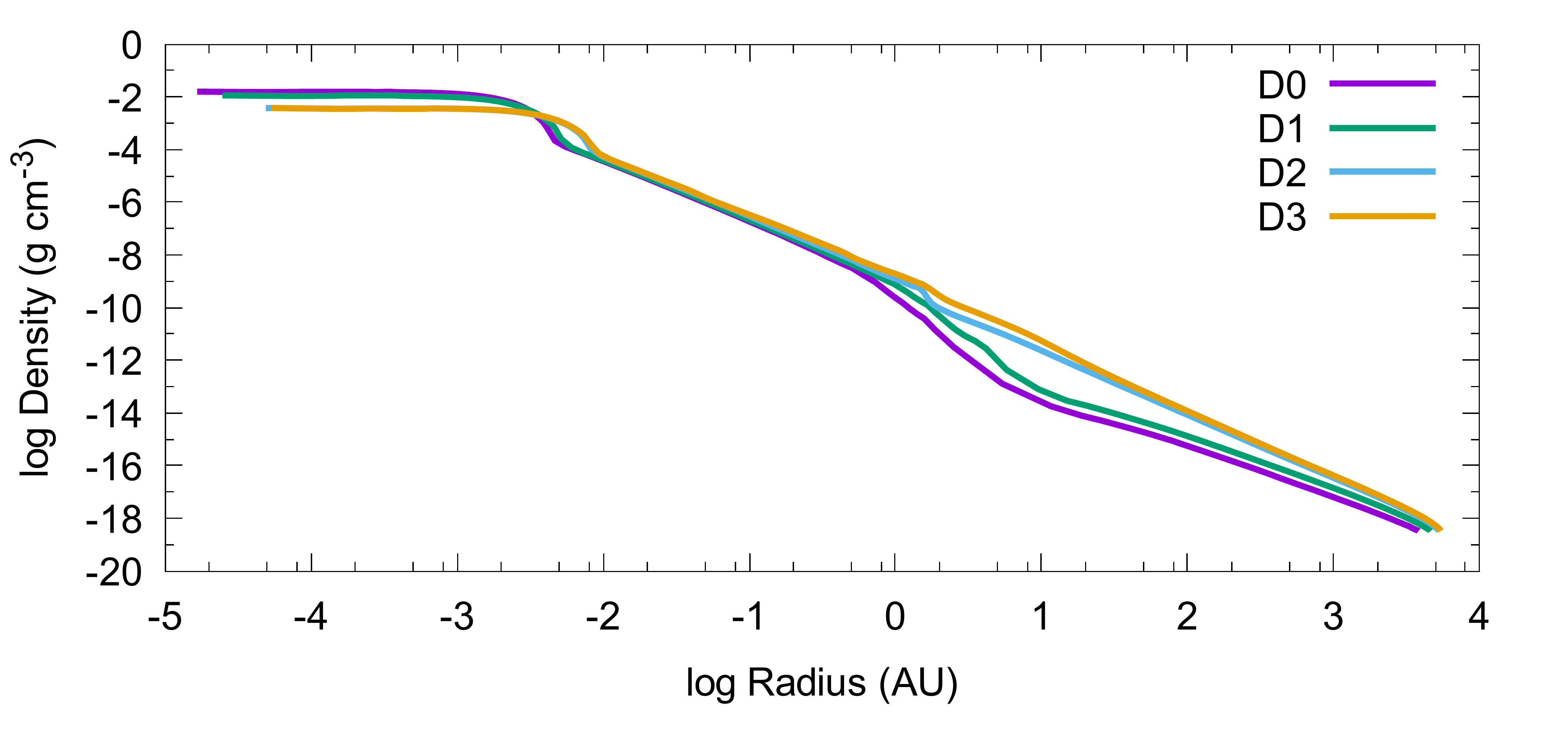}
\plotone{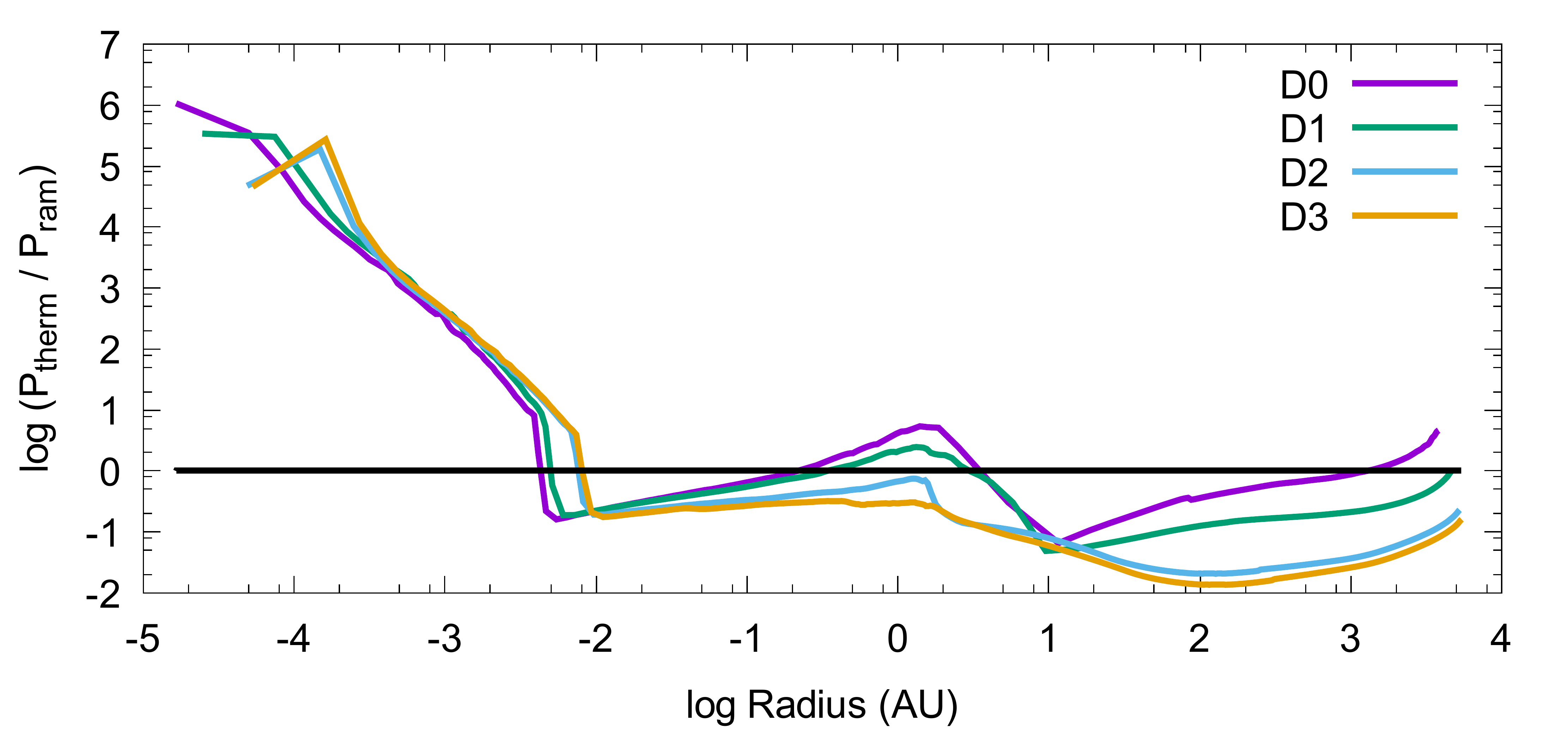}
\caption{\textit{Upper panel}: Final density profiles for runs D0 through D4, showing the transition to a collapse without the first core phase. \textit{Lower panel}: Ratio of thermal to ram pressure at the end of the simulation for the same four runs.}
\label{fig:fcless}
\end{figure}

We confirm this behavior in our runs D0 through D3 (Figure \ref{fig:fcless}). The initial conditions of D3 are identical to the ones \citet{Vaytet2017} used to investigate this phenomenon. The other three runs differ from D3 only in that their initial radii are smaller, which also implies a smaller mass and a smaller degree of instability. Figure \ref{fig:fcless} shows the transition between the cases with and without the first core. The first core's signature in the density profile, clearly visible at around 10 AU in runs D0 and D1, is absent in runs D2 and D3. The lower panel of Figure \ref{fig:fcless} confirms the importance of the ratio between thermal and ram pressure. In the first core-less cases, this ratio is smaller than unity everywhere outside the second core. The most favorable conditions for skipping the first core phase are those in which the ram pressure is very large (high degree of instability leading to large infall velocity), while the thermal pressure is relatively low.
With respect to the initial temperature of the cloud, there is one difference between our result and \citet{Vaytet2017}: They report that a collapse without the first core occurs only for an initial temperature of 5 K, but not for higher values such as 10 K or above, although our calculations show that it occurs even with 10 K. The boundary condition for all of our runs is calibrated such as to keep the initial temperature at 10 K, but the frequency-dependent radiative transfer then causes a decrease to $\approx$ 6 K (see section \ref{sec:CoreEvo}), so that the first core phase can be skipped despite the relatively high initial temperature. This example illustrates that, even though in most cases the overall behavior of gray and frequency-dependent calculations is similar, in certain cases such as this they may produce significantly different results. In summary, the unique condition for the formation of a protostar without the first core phase is that the initial core mass must be much larger than the Bonnor-Ebert mass (M $\geqslant$ 5 $\ix{M}{BE}$).

\section{Discussion}
\label{sec:Discussion}
\subsection{Spectral Energy Distribution}

\begin{figure}
\plotone{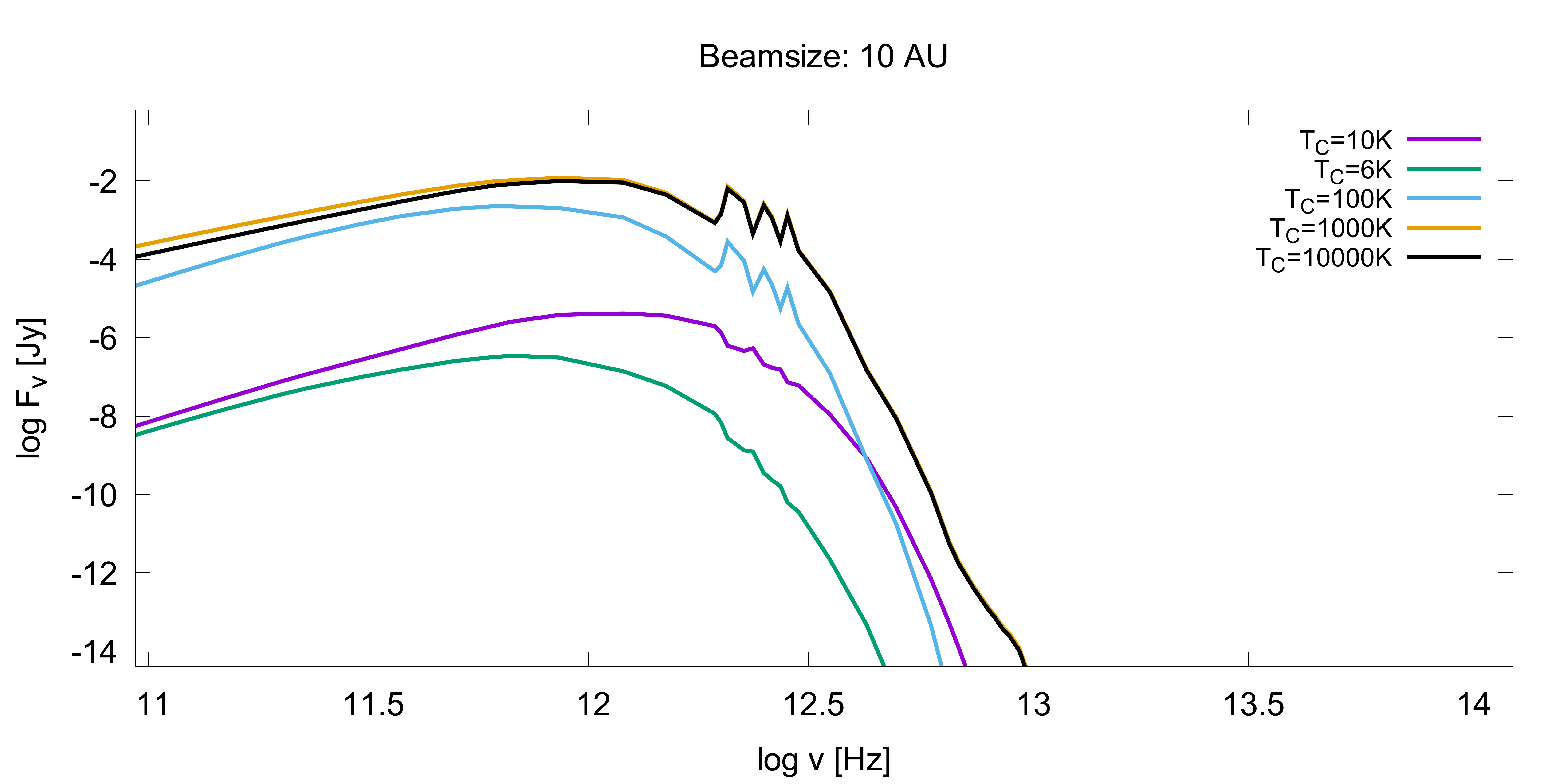}
\plotone{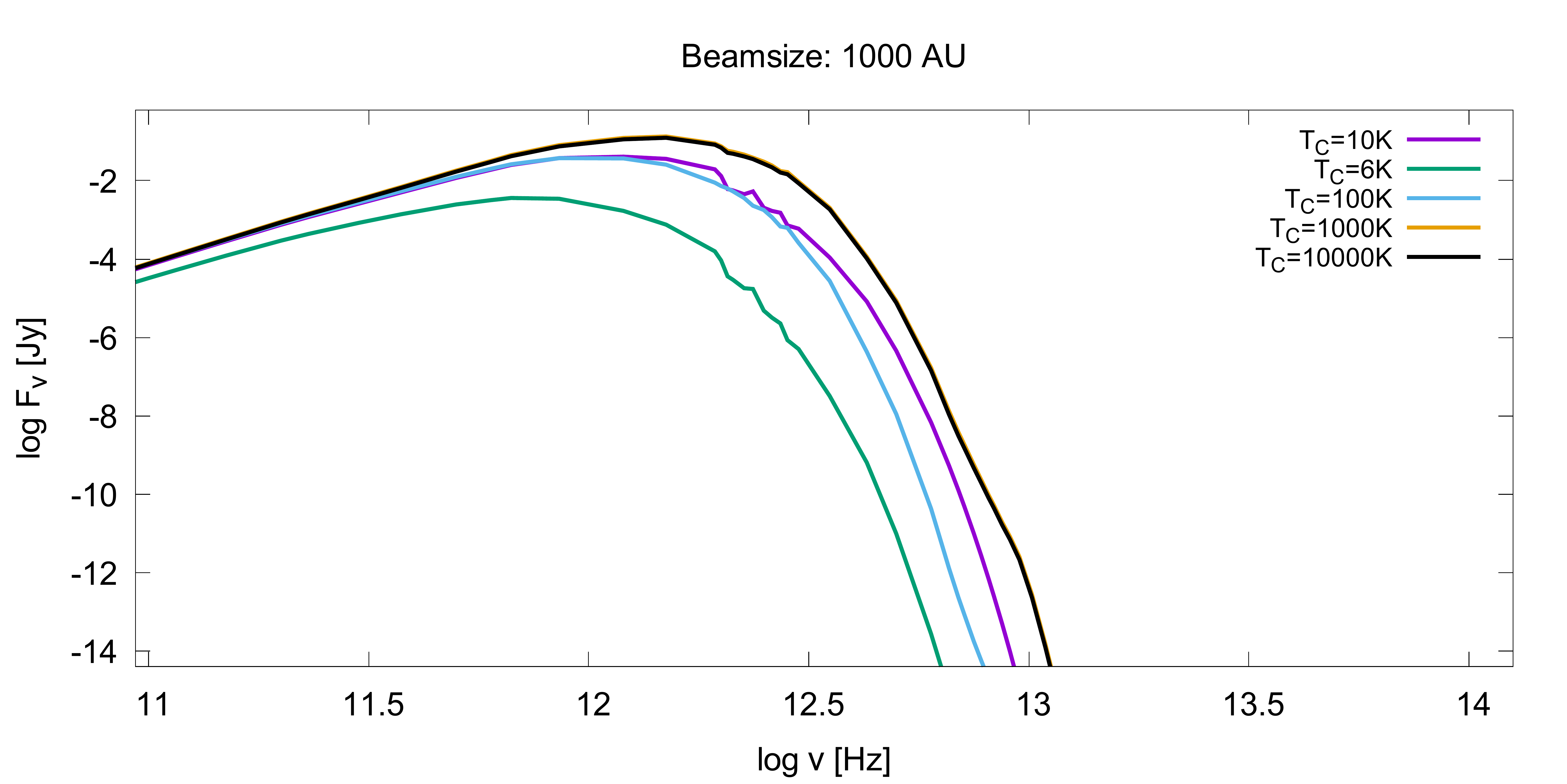}
\plotone{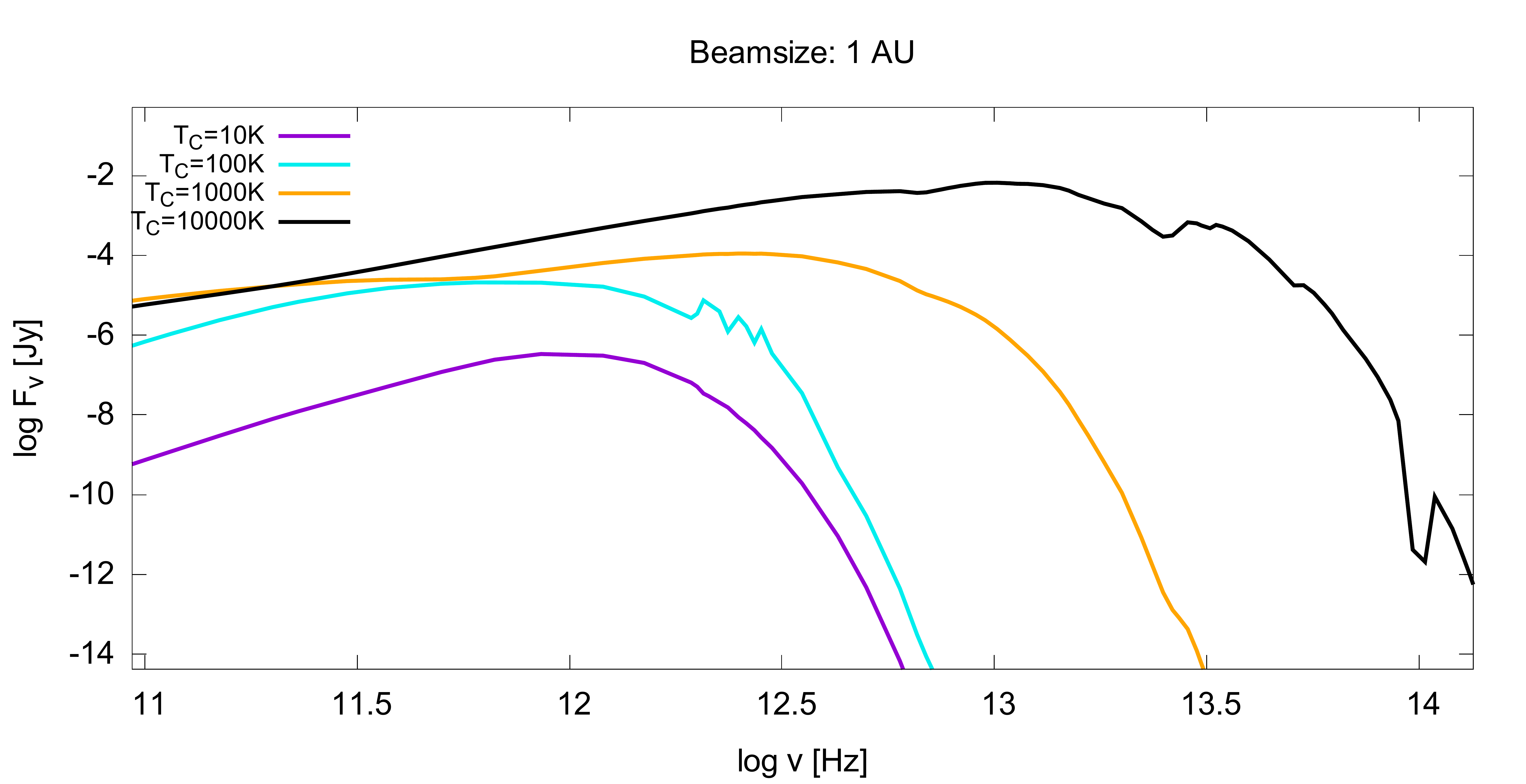}
\caption{Spectral energy distribution at various times for run B10 (upper two panels) and run B0 (lowest panel), at an assumed distance of 150 $pc$ and with different beamsizes. $T_{C}$ is the central temperature at the time of the snapshot. The 6K-plot is missing in the lower panel since run B0 never reached such low core temperatures. The evolution becomes clearer with smaller beamsizes, and is especially visible for the very low-mass cases where the obscuring envelope has disappeared.}
\label{fig:SED}
\end{figure}

We discuss the observational signature that would be expected from a brown dwarf in the process of formation. The total radiation escaping from the system is equal to what we call the ``boundary term'' in our radiative transfer scheme (\citet{Stamer2018}). By considering only a single direction for the escaping radiation, namely the direction towards the observer, we may use this to compute the spectral energy distribution (SED) that would be observed for a given beamsize at a given distance. For the distance, we assume 150 pc, and for the beamsize we vary between 1 and 1000 AU (Figure \ref{fig:SED}). For the calculation of the SEDs only, we increase the number of frequency bins to 400 for better frequency resolution. We note that for a relatively high-mass case such as B10, the emission is dominated by the cold envelope at all stages, i.e. the hot core region is obscured. This is also the main reason why observation of the early phases of star formation is difficult. The situation is exacerbated in the case of brown dwarfs because of their even lower luminosity compared to solar-type stars. A smaller beamsize helps to better distinguish different stages of the collapse, since the effect of the contribution of the envelope is reduced. These results essentially confirm those of \citet{inu98} and \citet{inu2000}, although the latter also found that the SED becomes markedly different in the following main accretion phase, which we have not considered. We do not show the SEDs of the A-runs or any other high mass B-runs since there are no significant differences. However, the situation is somewhat different for the very low-mass cases, as seen in the lowest panel of Figure \ref{fig:SED}. Here, the surrounding envelope has disappeared in the later stages, allowing the observer to look deeper into the warm interior. Therefore the SED is clearly warmer than at the beginning, and it is possible to look at a ``naked'' first core. 
Another difference is that for the very low mass case, the initial temperature decrease is less pronounced (run B0 never goes as low as 6K, unlike run B10). Recall that the temperature decrease is caused by blocking of the optical background radiation while low-frequency cooling radiation is still able to escape. Comparing runs B0 and B10, the initial density of B0 is larger by 2 orders of magnitude while the initial radius is smaller by about 1 order of magnitude. In consequence, B0's optical depth is about 1 order of magnitude larger, and in fact the initial state can be described as optically intermediate. As a result, the cloud becomes fully optically thick and the cooling radiation blocked as soon as the collapse begins, and therefore the central temperature hardly drops at all. \newline
Finally, we discuss the wavelike pattern that appears in most of the SEDs near $log$ $\nu = 12.4$. It stems from the \citet{Semenov2003} monochromatic opacity data, which shows this kind of pattern at the same frequency range. This opacity fine structure is only visible when viewing through an optically thin region, such as the initial cloud or, in later stages, the surrounding envelope. In the very low-mass case, it is mostly hidden in the later stages due to the lack of an envelope. It is even hidden in the initial state since in this run, the initial optical depth at the relevant frequencies is around unity.

\subsection{Observational Implications}

There are so far no confirmed observations of a first hydrostatic core, even though a number of candidate objects have been proposed. As a result of our simulations, we have found two effects that may affect future observations: First, the lifetime of the first core can be increased by about an order of magnitude for extremely low mass objects due to depletion of the envelope. Second, the lack of an obscuring envelope for these same objects exposes the first core and allows more high-frequency radiation to escape.
Both of these effects should make the observation of very low mass first cores easier than it would be otherwise. The longer lifetime increases the number of objects that exist, and the warmer SED makes them easier to identify and to separate them from more evolved objects. While their extremely low intrinsic luminosity makes them very hard to detect nonetheless, we propose that detection is somewhat less difficult than would be expected naively. Since our simulation is spherically symmetric and does not include the effects of rotation and magnetic fields, our quantitative results (i.e. the first core lifetime for a given mass) may not be very reliable. Qualitatively, however, we predict a bias towards very low mass objects in future confirmed observations of first cores.

\section{Conclusion}
\label{sec:Conclusion}

In this paper, we performed numerical simulations of spherically symmetric protostellar collapse, with a special focus on very low-mass objects, i.e. brown dwarfs. Our numerical model combines a second order Godunov hydrodynamics module with the frequency-dependent radiative transfer scheme we developed in a previous paper (\citet{Stamer2018}). We also include a realistic equation of state to account for the effects of hydrogen dissociation, as well as frequency-dependent opacities for dust and gas. Overall, this is still a very simplified model due to the spherical symmetry, which ignores multidimensional effects such as magnetohydrodynamics and rotation. Nevertheless, it is useful in investigating certain properties of the collapse process and of the objects formed as its result. 

When we were exploring numerical results with various initial conditions, we noticed a numerical problem with the hydrodynamics code that may be relevant to other investigations of self-similar gravitational collapse processes. To our knowledge, the solution for the problem has not been mentioned in the literature. Specifically, we found that in this self-gravitating system, a violation of energy conservation occurs if the effect of gravity in the energy equation is calculated according to the method of \citet{Colella1984}. Over time, this causes a spurious increase in the total energy, which in certain cases can blow the whole system apart due to the increased pressure. We provide a method to overcome this problem by explicitly calculating the gravitational energy in each shell before and after each timestep, as described in Appendix \ref{app:GravProb}.

Our collapse calculations showed that the properties of the protostar formed are mostly independent of the initial conditions of the cloud core, such as its mass and radius. This had previously been established for solar-mass and larger stars, but we confirmed that it also holds true at the low-mass end. The first core phase, however, showed some novel behavior. For the lowest-mass runs near the deuterium burning limit, the first core lifetime increases dramatically when lowering the mass and can exceed 10,000 years. This is due to the depletion of the object's gas envelope. At such low masses, essentially the entire cloud core's mass precipitates onto the first core, causing the accretion rate to become very small. This slows the rate of temperature increase inside the first core and therefore delays the onset of the second collapse, which requires a temperature of around 2000$K$.

This envelope depletion also has observational implications. We calculated the spectral energy distributions that would be observed from our simulations, and we found that the first core, which is usually obscured by the cold envelope, can be significantly more visible in the very low-mass cases where the envelope has disappeared. Combined with the much longer lifetime of these first cores, we conclude that very low mass first cores should be somewhat easier to observe than might be expected otherwise.

On the opposite end of the spectrum of first core lifetimes, we also confirmed that for sufficiently unstable setups, the first core stage may be skipped entirely, as has been previously shown by \citet{Vaytet2017}. A first core only forms if at some point the thermal pressure is larger than the ram pressure of the infalling gas. For very unstable and/or cold setups, this may never occur so the collapse directly proceeds to the formation of the second core. In contrast to \citet{Vaytet2017}, however, our simulations do not require a low initial temperature, since our frequency-dependent radiative transfer always causes the central temperature to drop to $\approx 6K$ due to blocking of the high-frequency stellar background radiation.

For our future work, we hope to expand these calculations, specifically our radiative transfer scheme, to multiple dimensions in order to obtain more realistic results. We are especially interested in the behavior of the first core lifetimes and their observational implications once effects such as rotation, magnetic fields, and outflows are included. In addition, with some modifications our code should also be applicable to the formation of planets through disk instability. Apart from the different boundary conditions that result from embedded in a protoplanetary disk, this is a situation which is very similar to protostellar collapse. 

\appendix
\section{Spurious Re-Expansion after Second Core Formation}
\label{app:GravProb}

\begin{figure}
\plotone{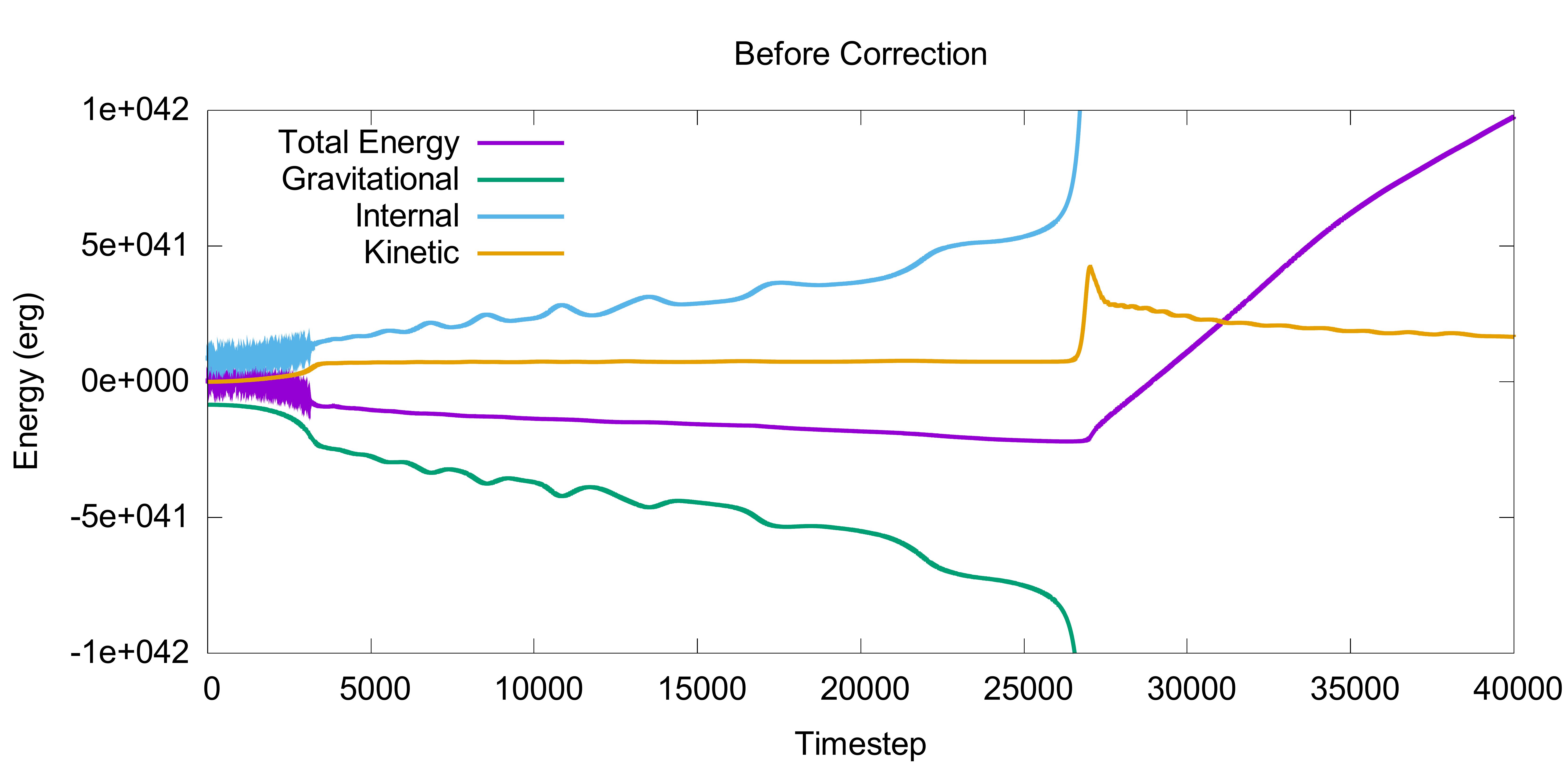}
\plotone{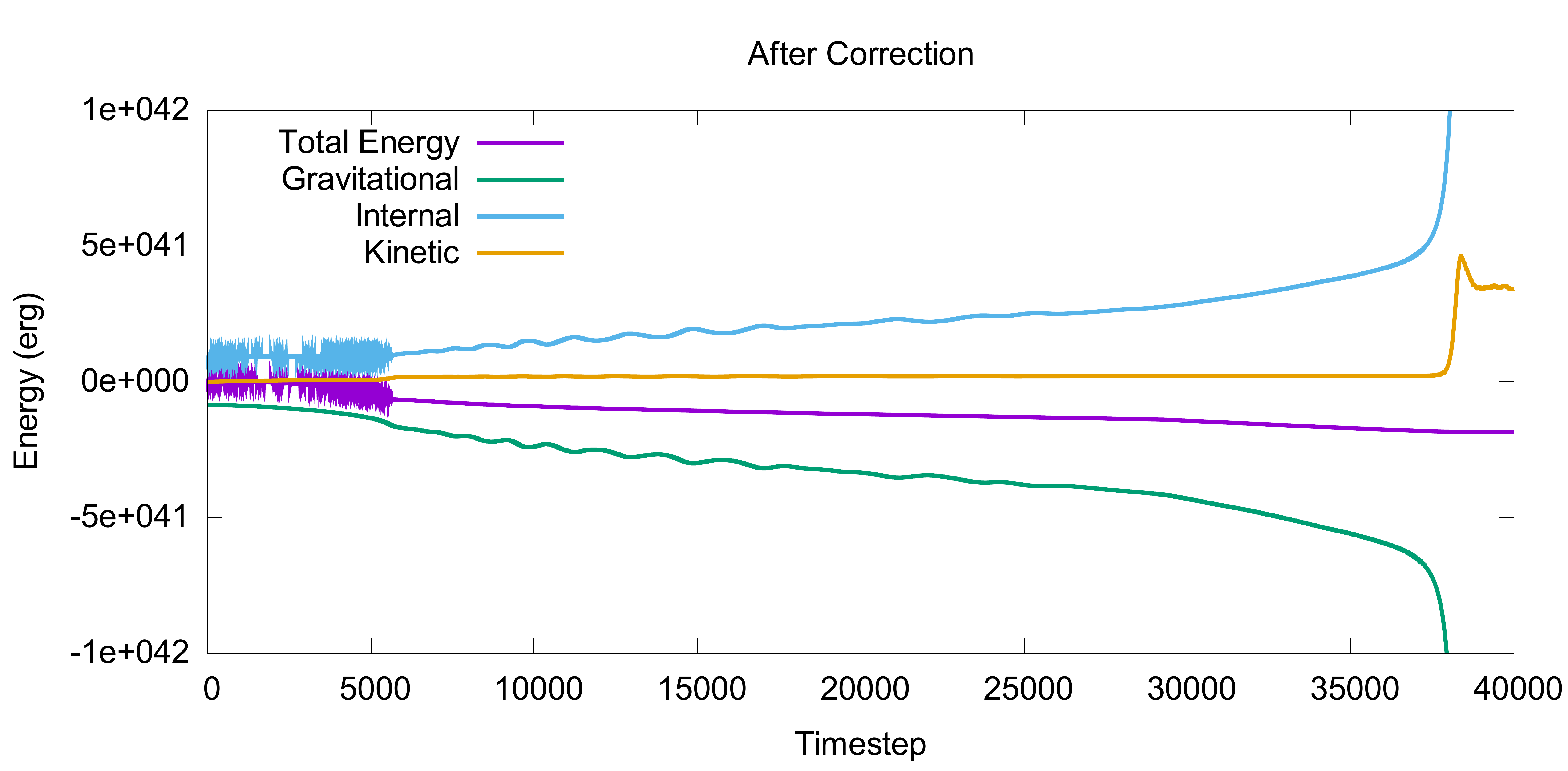}
\caption{Evolution of the system's total energy and its individual constituents in the original calculation (upper panel) and in the corrected version (lower panel). See text for details.}
\label{fig:EEvo}
\end{figure}

When we were testing our program, we encountered an unexpected behavior: After the formation of the second core, the system expanded again to first core densities and below. Suspecting a numerical error, we analyzed the development of the total energy, shown in Figure \ref{fig:EEvo}'s upper panel. In this plot, the formation of the first core takes place around timestep 5000, and that of the second core around timestep 27000. In the early phase, the internal energy oscillates rapidly as the system's energetics is dominated by the radiative exchange with the background. The magnitude of these oscillations depends on the size of the timestep. During the first core phase, we see a different kind of oscillation in the internal and gravitational energy. This is because the first core periodically expands and contracts slightly. During this time, the total energy slowly decreases due to radiation escaping from the system. The problem becomes apparent at the formation of the second core, when the total energy starts to increase steeply. This is obviously unphysical, since at this stage, radiative cooling should far exceed radiative and cosmic ray heating, and there are no other diabatic processes. In fact, this spurious increase occurred even when completely deactivating radiation and cosmic rays, implying that the cause is in the hydrodynamics module.

We found that the term $-v \nabla \Phi$ in equation \ref{eq:E} (the energy equation) was responsible for this error. This term represents the energy gained by the fluid due to conversion of gravitational potential energy. \citet{Colella1984} include a non-specified ``external body force'' $g$ in their scheme, in which case the numerical implementation of the resulting term $v g$ is an average of the value before the current timestep and the value after the timestep, i.e. $(v^{n} g^{n} + v^{n+1}g^{n+1}) / 2$. However, in our self-gravitating system, where $g=-\nabla \Phi$ is not constant, this approach is incorrect. The increase in specific energy when moving the fluid element in a gravitational field is $dU=-dr \nabla \Phi$. Taking the derivative with respect to time yields $dU/dt = -v \nabla \Phi$, but only if $\nabla \Phi$ is independent of time, which in our case it is not.

This inaccuracy leads to a continuous increase in the total energy of the system. In other words, the increase in the specific energy in each time step is larger than the corresponding decrease in gravitational potential energy. In the early stages of the collapse and during the first core phase, the effects of this error are too small to counteract the radiative losses, but during second core formation much more gravitational energy is converted to internal energy, and the abnormally large increase in pressure ultimately blows apart the second core. 

We have solved this problem by explicitly calculating the gravitational energy in each shell before and after the time step, and inserting the difference in the energy equation in place of the term $-v \nabla \Phi$. The gravitational potential energy of any shell is given by

\begin{equation}
\ix{E}{grav}=-G \int_{\ix{m}{I}}^{\ix{m}{O}} \frac{m} {r(m)} dm,
\end{equation}

where $m$ is equal to the enclosed mass. $\ix{m}{I}$ and $\ix{m}{O}$ are the values of this mass coordinate at the inner and outer boundary of the shell. Expressing $m$ as a function of the radial coordinate $r$, we obtain

\begin{equation}
\ix{E}{grav}=-G \int_{\ix{r}{I}}^{\ix{r}{O}} \frac{\ix{m}{I} + \frac{4 \pi}{3} \rho \left(r^3 - \ix{r}{I}^3 \right) }{r} 4 \pi \rho r^2 dr.
\end{equation}

The integral in the above equation can be solved analytically:

\begin{align}
\begin{split}
&4 \pi \rho \int \left( \ix{m}{I} + \frac{4 \pi}{3} \rho \left(r^3 - \ix{r}{I}^3 \right) \right) r dr \\
&=4 \pi \rho \left( \left( \frac{\ix{m}{I}}{2}  - \frac{2 \pi}{3} \rho \ix{r}{I}^3 \right) r^2 + \frac{4 \pi}{15} \rho r^5 \right) + C.
\end{split}
\end{align}

Since all necessary values are known both before and after the timestep, the change in gravitational energy can be calculated and inserted into the energy equation. 

The lower panel of Figure \ref{fig:EEvo} shows the behavior after applying this correction. The main difference is that there is no longer any increase in total energy at the point of second core formation, which here occurs around timestep 37000. In addition, the oscillations during the first core phase are significantly smaller.

\begin{figure}
\plotone{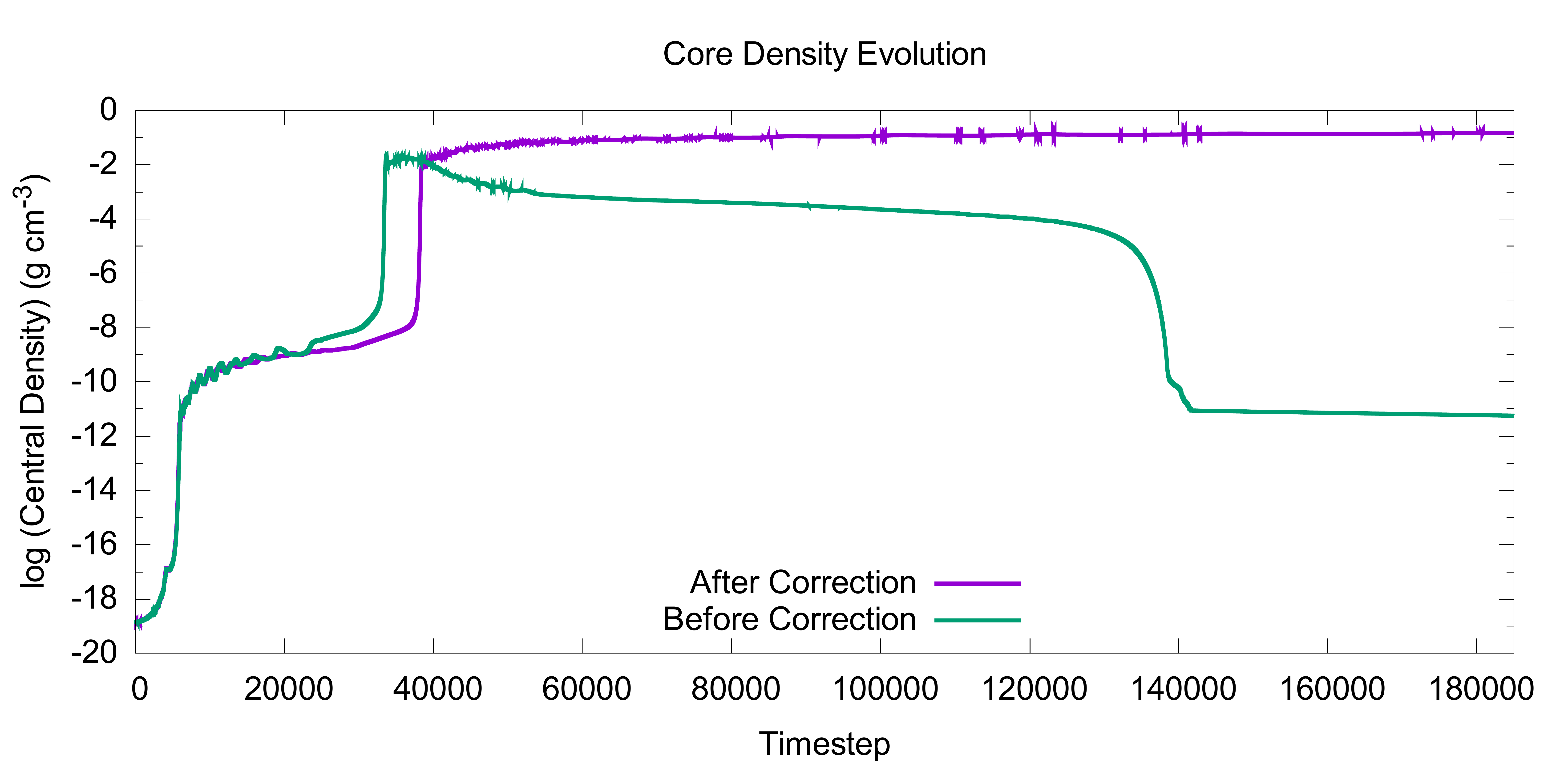}
\caption{Evolution of the central density in the original calculation and in the corrected version. Only the former shows the unexpected re-expansion and destruction of the second core.}
\label{fig:RhoEvoEnergetics.pdf}
\end{figure}

Figure \ref{fig:RhoEvoEnergetics.pdf} shows the evolution of the central density before and after the correction. In the improved version, the second core is stable and no unphysical re-expansion occurs.

\end{document}